\documentclass[useAMS,usenatbib]{mnras}
\usepackage{graphicx,natbib,times,mathptm,bm,amssymb,enumitem}
\usepackage[fleqn]{amsmath}
\def\aap{A\&A}

\def\apj{ApJ}

\def\mnras{MNRAS}

{\newif\ifnotend
\notendtrue
\def\veclist{ABCDEFGHIJKLMNOPQRSTUVWXYZabcdefghijklmnopqrstuvwxyz.}
\def\top#1#2.{#1}
\def\tail#1#2.{#2.}
\loop\expandafter\xdef\csname v\expandafter\top\veclist\endcsname%
{{\noexpand\bm\expandafter\top\veclist}}
\edef\veclist{\expandafter\tail\veclist}
\if\veclist.\notendfalse\fi\ifnotend\repeat}
\let\boldgrk=\gkvecten
\let\boldgrksc=\gkvecseven
\def\gkthing#1{{\mathchoice%
        {\hbox{{\boldgrk\char#1}}}
        {\hbox{{\boldgrk\char#1}}}
        {\hbox{{\boldgrksc\char#1}}}
        {\hbox{{\boldgrksc\char#1}}}}}

\def\vtheta{\gkthing{18}}

\def\vSigma{\bm{\Sigma}}

\def\d{{\rm d}}

\def\pr{{\mathop{\hbox{p}}}}

\def\aap{A\&A}

\def\apj{ApJ}

\def\mnras{MNRAS}

\def\app#1#2{%
  \mathrel{%
    \setbox0=\hbox{$#1\sim$}%
    \setbox2=\hbox{%
      \rlap{\hbox{$#1\propto$}}%
      \lower1.1\ht0\box0%
    }%
    \raise0.25\ht2\box2%
  }%
}
\def\appropto{\mathpalette\app\relax}

\bibpunct{(}{)}{;}{a}{}{,}

\def\macromodel{\Theta}
\def\micromodel{\zeta}
\def\galaxymodel{\beta}

\def\vytilde{\tilde{\bm{y}}}
\def\vltilde{\tilde{\bm{l}}}
\def\vbtilde{\tilde{\bm{b}}}
\def\vatilde{\tilde{\bm{a}}}
\def\vstilde{\tilde{\bm{s}}}

\def\hyper{\bm{\theta}}

\bibpunct{(}{)}{;}{a}{}{,}

\begin{document}

\bibliographystyle{mnras}

\title[Large scale 3d GP extinction mapping]{Large-scale three-dimensional Gaussian process extinction mapping}

\author[Sale \& Magorrian]{S.~E.~Sale$^{1}$, and J.~Magorrian$^2$\\
$^1$ Astrophysics Group, School of Physics, University of Exeter, Stocker Road, Exeter EX4 4QL, UK\\
$^2$ Rudolf Peierls Centre for Theoretical Physics, Keble Road, Oxford OX1 3NP, UK\\}

\date{Received .........., Accepted...........}

\maketitle

\begin{abstract}

\noindent Gaussian processes are the ideal tool for modelling the Galactic ISM, combining statistical flexibility with a good match to the underlying physics.
In an earlier paper we outlined how they can be employed to construct three-dimensional maps of dust extinction from stellar surveys.
Gaussian processes scale poorly to large datasets though, which put the analysis of realistic catalogues out of reach.
Here we show how a novel combination of the Expectation Propagation method and certain sparse matrix approximations can be used to accelerate the dust mapping problem.
We demonstrate, using simulated Gaia data, that the resultant algorithm is fast, accurate and precise.
Critically, it can be scaled up to map the Gaia catalogue. 

\end{abstract}

\section{Introduction}\label{sec:intro}

Our Galaxy's stellar content is the focus of many ambitious, large-scale surveys, the most notable of these being Gaia \citep{Gaia_Prusti.2016}.
The interpretation of the resulting stellar catalogues is hampered by the presence of interstellar dust, which becomes particularly severe near the Galactic plane.
Any models for the structure and evolution of the Galaxy constructed from these data must therefore account for the three-dimensional distribution of extincting dust.  On the other hand, dust is an important tracer of the densest parts of the Galaxy's ISM.  Therefore the three-dimensional dust distribution is of interest in its own right.

In \citet[][hereafter Paper~I]{Sale_Magorrian.2014} we developed a method for mapping interstellar extinction in three dimensions by modelling the (logarithm of the) extinction as a Gaussian process\footnote{Gaussian Processes on two or three dimensional spaces are sometimes referred to as `Gaussian Random Fields', indeed this was the terminology we employed in Paper I.} (hereafter GP).
Paper~I builds on earlier work by \cite{Vergely_FreireFerrero.2001} and overcomes the most blatant shortcomings encountered in previous work, such as \cite{Marshall_Robin.2006}, \cite{Chen_Schultheis.2013}, \cite{Sale_Drew.2014} and \cite{Green_Schlafly.2015}:
it banishes the `fingers of God' that plague other 3D extinction maps; the treatment of small scale variations in extinction is much improved, enabling the extinction to a particular point in space to be estimated far more precisely than previously possible.
Subsequently, \cite{RezaeiKh._Bailer-Jones.2017} have described a very similar method.

Modelling the logarithm of extinction as a GP results in a simple mathematical model that exhibits compelling parallels to the true ISM.
The extinction to any given point (or set of points) in space is then described by a (multivariate) lognormal distribution, consistent with theoretical arguments \citep{Ostriker_Stone.2001, Nordlund_Padoan.1999} that the integrated extinction to any given point is approximately proportional to the product of many small, independent relative compressions or rarefactions of the underlying density field along the line of sight to that point.
In any GP the statistical properties of the fluctuations about the mean are completely defined by the covariance function.
In our formulation the covariance function is directly related to the power spectrum of density fluctuations in the ISM.
Our model includes a physical treatment of interstellar turbulence, such as that given by \cite{Kolmogorov_only.1941}, that provides us with a form for the power spectrum of the ISM.
This turbulent power spectrum is truncated at a spatial scale that corresponds to the energy injection scale in the ISM.

In Paper~I we identified some practical challenges in applying our method to the construction of large-scale extinction maps.
The purpose of the present paper is to demonstrate how these challenges can be overcome.
We begin in section~\ref{sec:GPs} by reviewing the topic of GP Regression.
We explain how we use it to map extinction, pointing out the practical difficulties that arise in section~\ref{sec:mappingasGP}.
In section~\ref{sec:scheme} we describe a scheme that uses a combination of two approximate algorithms that make the production of large-scale extinction maps feasible.
We verify the validity of our implementation in section~\ref{sec:verify} by using it to reconstruct a 3D extinction map from simulated Gaia data, before summing up in Section~\ref{sec:close}.

\section{Gaussian Process regression}\label{sec:GPs}

We model (the logarithm of) the extinction to points in three-dimensional space
as a GP.
As for any GP, the statistical properties of this
density field are completely described by just two functions: a
function $m(x)$ that gives the expectation value of the field at any
position~$x$, and a function $\Sigma(x_1,x_2)$ that returns the covariance
of the field values between any pair of points $(x_1,x_2)$.
Our method for mapping extinction in essence reduces to a GP regression problem.
In this section we provide a brief review of GPs and how they are employed in regression problems.
Readers interested in a more comprehensive introduction are directed to \cite{Rasmussen_Williams.2005}.

The defining property of a GP is that, for any set
of $N\ge1$ points $\vx=\{x_1,\ldots,x_N\}$ in this
three-dimensional space, the joint PDF of the values
$\vf=\{f_1,\ldots,f_N\}$ of the GP at those points is Gaussian,
namely
\begin{equation}
  \vf\sim \mathcal{N}(\vm, \vSigma),
  \label{eq:GPdefn}
\end{equation}
where $\mathcal{N}$ indicates the multivariate Gaussian distribution,
in which the mean vector $\vm(\vx)$ and covariance matrix $\vSigma$ have
elements $\vm_n=m(x_n)$ and $\vSigma_{ij} = \Sigma(x_i, x_j)$
respectively.
 
Now suppose that we know the values $\vf$ of the field at some points $\vx$, but want to infer values $\vf_\star$ at some other points $\vx_\star$.
From~\eqref{eq:GPdefn} it immediately follows that the joint distribution of $\vf$ and $\vf_\star$ is
\begin{equation}
\begin{bmatrix} 
\vf \\ 
\vf_{\star} 
\end{bmatrix} \sim \mathcal{N} \left(
\begin{bmatrix}
\vm(\vx) \\
\vm(\vx_{\star} )
\end{bmatrix} , 
\begin{bmatrix}
\vSigma(\vx,\vx) & \vSigma(\vx,\vx_\star) \\
\vSigma(\vx,\vx_\star)^{\rm T} & \vSigma(\vx_\star,\vx_\star)
\end{bmatrix} \right),
\end{equation}
in which the sub-matrices $\vSigma(\va,\vb)$ have elements
$[\vSigma(\va,\vb)]_{ij}=\Sigma(a_i,b_j)$ and so on.
Then, using standard properties of Gaussians
\citep[e.g.,][]{Bishop_only.2006}, the conditional distribution of the
$\vf_\star$ is another Gaussian
\begin{equation}
\vf_{\star}|\vf \sim \mathcal{N} ( \vm_{\vf_\star|\vf},\vSigma_{\vf_\star|\vf})
\end{equation}
whose mean and covariance are given by
\begin{equation}
  \begin{split}
    \vm_{\vf_\star|\vf} &= \vm(\vx_\star)+\vSigma^{\rm T}(\vx,\vx_\star)\vSigma^{-1}(\vx,\vx)(\vf-\vm(\vx)),\\
    \vSigma_{\vf_\star|\vf} &=
    \vSigma(\vx_\star,\vx_\star)-\vSigma^{\rm T}(\vx,\vx_\star)\vSigma^{-1}(\vx,\vx)\vSigma(\vx,\vx_\star).
\end{split}
\end{equation}
Therefore, with some simple matrix algebra, we can estimate the values $\vf_\star$ of the field at the
locations $\vx_\star$ given (perfect) observations of it elsewhere.

This result is easily generalised to the situation in which
our observations $\widetilde\vf$ of the underlying values $\vf$ are
subject to some measurement error.
Suppose that the measurement errors for each point are Gaussian and
mutually independent, namely
\begin{equation}
\widetilde f_n = f_n + \epsilon_n, \quad {\rm where} \quad \epsilon_n \sim \mathcal{N} (0, \sigma_n),
\end{equation}
so that
\begin{equation}
\widetilde\vf \sim \mathcal{N}\left(\vm(\vx), \vSigma(\vx,\vx)+\vSigma_{\rm m}\right), \label{eqn:GP+noise}
\end{equation}
where $\vSigma_{\rm m}$ is a diagonal matrix with the observational
uncertainties $\sigma_n$ along the diagonal.  
Then the joint distribution of $\widetilde\vf$ and $\vf_\star$ is
\begin{equation}
\begin{bmatrix} 
\widetilde\vf \\ 
\vf_{\star} 
\end{bmatrix} \sim \mathcal{N} \left(
\begin{bmatrix}
\vm(\vx) \\
\vm(\vx_{\star} )
\end{bmatrix} , 
\begin{bmatrix}
\vSigma(\vx,\vx)+\vSigma_{\rm m} & \vSigma(\vx,\vx_\star) \\
\vSigma(\vx,\vx_\star)^{\rm T} & \vSigma(\vx_\star,\vx_\star)
\end{bmatrix} \right),
\end{equation}
from which it follows that the posterior distribution of the $\vf_\star$ is again a Gaussian
\begin{equation}
\vf_{\star}|\widetilde\vf \sim \mathcal{N} ( \vm_{\vf_\star|\widetilde\vf},\vSigma_{\vf_\star|\widetilde\vf})
\label{eqn:cond_GP}
\end{equation}
whose mean and covariance are given by
\begin{equation}
  \begin{split}
    \vm_{\vf_\star|\widetilde\vf} &= \vm(\vx_\star)+\vSigma^{\rm
      T}(\vx,\vx_\star)[\vSigma(\vx,\vx)+\vSigma_{\rm m}]^{-1}(\widetilde\vf-\vm(\vx)),\\
    \vSigma_{\vf_\star|\widetilde\vf} &=
    \vSigma(\vx_\star,\vx_\star)-\vSigma^{\rm
      T}(\vx,\vx_\star)[\vSigma(\vx,\vx)+\vSigma_{\rm m}]^{-1}\vSigma(\vx,\vx_\star).
\end{split}
\label{eqn:cond_GP0}
\end{equation}
So, we can estimate the values $\vf_\star$ of the field at the
locations $\vx_\star$ given noisy observations of it elsewhere.
This is known as Gaussian process regression.
GP regression is a practical solution in many regression problems since it does not require one to assume some simple parametrized form for the data.

The implementation of GP regression can typically be divided into two stages.
First there is learning/training phase in which inferences are made on
any required latent or unknown parameters, especially those that
define the mean and/or covariance functions, $m(x)$ and~$\Sigma(x,x')$.
Then there is the prediction phase in which equations
\eqref{eqn:cond_GP} and~\eqref{eqn:cond_GP0} are used to predict of the values of the
process for locations where observations are not available. 

Consider the situation where the hyperparameters, $\vtheta$, that set the mean and covariance functions, are unknown.
Then, in the learning phase, we seek to determine the posterior distribution of the hyperparameters given the observations $\widetilde\vf$, 
\begin{equation}
\pr(\vtheta | \widetilde\vf) = \frac{\pr(\widetilde\vf | \vtheta) \pr(\vtheta)}{\int \, d\vtheta \, \pr(\widetilde\vf | \vtheta) \pr(\vtheta)} \label{eqn:example_learning}
\end{equation}
where $\pr(\widetilde\vf | \vtheta)$ is a multivariate Gaussian PDF, as in~\eqref{eqn:GP+noise} and $\pr(\vtheta)$ some hyperprior.
Then, as the prediction phase, this can be combined with the expression~\eqref{eqn:cond_GP} for $\pr(\vf_{\star}|\widetilde\vf, \vtheta)$ to give a posterior on $\vf_{\star}$:
\begin{equation}
\pr(\vf_{\star}|\widetilde\vf) = \int \, d\vtheta \, \pr(\vf_{\star}|\widetilde\vf, \vtheta)  \pr(\vtheta | \widetilde\vf) . \label{eqn:example_prediction}
\end{equation}
With all but the smallest datasets the computational cost of
calculating of the probabilities in the learning phase of GP
regression is dominated by solving for the inverse of the covariance
matrix, $[\vSigma(\vx,\vx)+\vSigma_{\rm m}]^{-1}$ that appears in~\eqref{eqn:cond_GP0}.
Given $N$ observations the CPU time cost of solving for the inverse of the $N \times N$ matrix scales as $\mathcal{O}(N^3)$ and memory requirements as $\mathcal{O}(N^2)$.

\section{Extinction mapping as  a Gaussian process regression problem}
\label{sec:mappingasGP}

\begin{table*}
\begin{tabular}{c|l}
$(\tilde l_i, \tilde b_i)$ & the $(l,b)$ coordinates observed for the $i^{\rm th}$ star in the catalogue.\\
$\tilde{y}_i$ & all other directly observed quantities (e.g., broad-band fluxes, trigonometric parallax)
for the $i^{\rm th}$ star.\\
\noalign{\smallskip}
$s_i$ & the distance to the $i^{\rm th}$ star in the catalogue. \\
$A_i$ & the extinction to the $i^{\rm th}$ star in the catalogue. \\
$a_i$ & the logarithm of extinction to the $i^{\rm th}$ star in the catalogue, i.e. $a_i = \ln A_i$. \\
$z_i$ & the component of the Gaussian mixture model used to describe the likelihood $\pr(\tilde y_i,s_i,z_i|a_i,\tilde l_i,\tilde b_i,\micromodel,\galaxymodel)$. \\
\noalign{\smallskip}
$\macromodel$ & parameters that set the large-scale distribution of extinction.\\
$\micromodel$ & the model for the small scale turbulent variations of extinction.\\
$\galaxymodel$ & the background Galaxy model, including prior on position, metallicity, age, etc. \\
\end{tabular}
\caption{A list of the parameters adopted in our model.
\label{tab:notation} }
\end{table*}

As in Paper~I, we assume that $a(x)$, the logarithm of the 
extinction to the point~$x$, is described by a GP.  Then the joint
PDF $\pr(\va|\vl,\vb,\vs,\macromodel,\micromodel)$ of the log extinctions to many points is a
multivariate Gaussian.
The parameters $\macromodel$ control how the mean extinction varies with position $x=(l,b,s)$.  
We demonstrated in Paper~I how to derive the covariance function of extinction.
This recognised that the 3d extinction map is a projection of the 3d density map, so that the extinction covariance function is obtained by integrating the density covariance along (any) two sightlines. 
As a result, the (extinction) covariance function depends on the the small-scale turbulent structure of the ISM ($\micromodel$) and $\macromodel$.
Table~\ref{tab:notation} provides a summary of our notation.

The remaining important ingredients of our model include a Galaxy model $\galaxymodel$ that describes the distribution of the different stellar populations within the Galaxy and of the underlying stellar physics.
This supplies us with a likelihood function $\pr(\tilde y|\tilde l, \tilde b,s,a,\galaxymodel)$ for the stellar observables $\tilde y$ (such as apparent magnitudes, effective temperature) that depend on the distance $s$ and log extinction~$a$ to the star, as well as the star's position within the galaxy (because the mixture of stellar populations changes with position).
We assume that the observations, $\tilde l$ and $\tilde b$, of the Galactic
coordinates of each star are measured exactly, so that we can
condition all the following analysis on these values\footnote{Formally
  we assume the likelihood on the true Galactic coordinates of each
  star can be approximated by a delta-function and then marginalise
  the true Galactic coordinates.}.

Therefore the probability of anything in our model can be obtained
simply by marginalising the innocuous-looking
\begin{equation}
  \begin{split}
&  \pr(\vltilde,\vbtilde,\vs,\vytilde,\va,\macromodel,\micromodel,\galaxymodel)\\
&\quad=\pr(\galaxymodel)\pr(\micromodel)\pr(\macromodel|\micromodel)
\pr(\vltilde,\vbtilde,\vs|\galaxymodel)
\pr(\va|\vltilde,\vbtilde,\vs,\macromodel,\micromodel)
\pr(\vytilde|\vltilde,\vbtilde,\vs,\va,\galaxymodel),
  \end{split}
\label{eqn:peverything0}
\end{equation}
in which the first three factors are our model's priors.
The third, $\pr(\vltilde,\vbtilde,\vs|\galaxymodel)$, is the spatial
density of {\it all} stars in our assumed Galaxy model~$\galaxymodel$ and provides a prior on the position of each observed star.
The last two factors are more challenging.
As discussed above,
$\pr(\va|\vltilde,\vbtilde,\vs,\macromodel,\micromodel)$ is just a
multivariate Gaussian, the only difficulty being the sheer size of its
covariance matrix.  The final factor
\begin{equation}
\pr(\vytilde|\vltilde,\vbtilde,\vs,\va,\micromodel,\galaxymodel)
=\prod_{n=1}^N  
\pr(\tilde y_n|\tilde l_n,\tilde b_n,s_n,a_n,\micromodel,\galaxymodel)
\label{eqn:likprod}
\end{equation}
relates each star's observables~$y_n$ to its position within the
Galaxy.

We can simplify~\eqref{eqn:likprod} at the cost of introducing a new, latent variable.
In \cite{Sale_Magorrian.2015} we showed how the
product $\pr(\tilde y|\tilde l,\tilde b,s,a,\galaxymodel)\pr(s|\tilde l,\tilde b,\galaxymodel)$ could be approximated as a mixture of
Gaussians in $(a,\log s)$.  Each Gaussian component might be identifiable with, e.g., a distinct stellar luminosity class, but we emphasise that this Gaussian-mixture approximation is introduced primarily for computational convenience.
Let us introduce a
latent variable $z$, which is a vector that indicates which Gaussian
the star was drawn from: if there are $M$ Gaussians in the mix
representing $\pr(\tilde y|\tilde l,\tilde b,s,a,\galaxymodel)\pr(s|\tilde l,\tilde b,\galaxymodel)$ then
$z$ has $M$ elements, just one of which is equal to~1, the rest being zero.
Then
\begin{equation}
  \pr(\tilde y,s,z|a,\tilde l,\tilde b,\micromodel,\galaxymodel)
= 
\prod_{m=1}^M\left[G_m(\log s,a)\right]^{z_m},
\label{eqn:gmm_approx}
\end{equation}
where $G_m(\log s,a)$ is the $m^{\rm th}$ Gaussian, whose mass, mean
and covariance depend on $(\tilde y, \tilde l,\tilde b)$. 

With the latent variable~$\vz$, the probability of everything in our model becomes
\begin{equation}
  \begin{split}
&  \pr(\vltilde,\vbtilde,\vs,\vytilde,\va,\vz,\macromodel,\micromodel,\galaxymodel)\\
&\quad=\pr(\galaxymodel)\pr(\micromodel)\pr(\vltilde,\vbtilde|\galaxymodel)
\pr(\macromodel|\micromodel)\pr(\va|\vltilde,\vbtilde,\vs,\macromodel,\micromodel)
\pr(\vytilde,\vs,\vz|\vltilde,\vbtilde,\va,\micromodel,\galaxymodel),
\end{split}
\label{eqn:peverything}
\end{equation}
where
\begin{equation}
\pr(\vytilde,\vs,\vz|\vltilde,\vbtilde,\va,\micromodel,\galaxymodel)= 
\prod_{n=1}^N\prod_{m=1}^{M_n}[G_{nm}(\log s_n,a_n)]^{z_{nm}}.
\label{eqn:vectormultiGaussian}
\end{equation}
Marginalising the latent variable~$\vz$ in these expressions gives the
more physically meaningful~\eqref{eqn:peverything0}, hiding the
details of our Gaussian mixture parametrization.

\subsection{What do we want to know?}
As discussed in Paper~I, there are four broad tasks contained under the umbrella of extinction mapping:
\begin{enumerate}[leftmargin=*,labelindent=0em]
\item inferring the extinction and/or distance to stars within an observed catalogue;
\item inferring the extinction to some arbitrary locations;
\item determining the large-scale distribution of extinction and, by
  extension, dust; and
\item constraining the turbulent physics of the ISM by studying
  how extinction varies on small scales.
\end{enumerate}
All of these tasks are solved by marginalising~\eqref{eqn:peverything} appropriately.
For example, finding distances and extinctions of the stars within the
catalogue, task (i), involves the posterior
\begin{equation}
  \begin{split}
  \pr(\vs,\va,\macromodel|\vytilde,\vltilde,\vbtilde,\micromodel,\galaxymodel) 
&= \frac1{\pr(\vytilde,\vltilde,\vbtilde,\micromodel,\galaxymodel)}
{\sum_{\vz}\pr(\vltilde,\vbtilde,\vs,\vytilde,\va,\vz,\macromodel,\micromodel,\galaxymodel)},
  \end{split}
\label{eqn:posterior}
\end{equation}
where the evidence $\pr(\vltilde,\vbtilde,\vytilde,\micromodel,\galaxymodel)$ in the denominator, which
depends only on directly observed variables, is an
uninteresting normalisation constant unless one wants to compare
different assumptions for~$\micromodel$ or~$\galaxymodel$.
Marginalising $\va$ and $\vs$ from
this gives the posterior distribution of the large-scale distribution of
dust~$\macromodel$,
\begin{equation}
  \begin{split}
 &
 \pr(\macromodel|\vytilde,\vltilde,\vbtilde,\micromodel,\galaxymodel) 
\\
&\quad= \frac1{\pr(\vytilde,\vltilde,\vbtilde,\micromodel,\galaxymodel)}
{\sum_{\vz}\int\d\vs\int\d\va\,\pr(\vltilde,\vbtilde,\vs,\vytilde,\va,\vz,\macromodel,\micromodel,\galaxymodel)},
  \end{split}
\label{eqn:posteriormacro}
\end{equation}
which solves task (iii).  
Given the Gaussian mixture description we adopt from
\cite{Sale_Magorrian.2015}, the inner integral over $\va$, for some
$\vz$ and $\vs$ can be found trivially.
Similarly, multiplying \eqref{eqn:posteriormacro} by the prior
$\pr(\micromodel)$ and marginalising~$\macromodel$ gives the posterior
distribution of~$\micromodel$.  
For the time being, however, we will ignore the possibility of examining the
small-scale characteristics of dust, although we intend to return to
this issue in a future paper.

Determining the distributions above essentially constitutes the learning phase of the GP regression.
Subsequently in the prediction phase of the GP regression we can estimate the extinction to some arbitrary points in space, i.e. task (ii).
If performed on some regular grid of locations this constitutes the production of an extinction map.
We can extend the probability of everything~\eqref{eqn:peverything} to
include the PDF of extinctions~$\va_\star$ to given set of locations
$\vx_\star\equiv(\vl_{\star}, \vb_{\star},\vs_{\star})$.  The result is
\begin{equation}
  \begin{split}
&  \pr(\vltilde,\vbtilde,\vs,\vytilde,\va,\va_\star,\vz,\macromodel,\micromodel,\galaxymodel|\vx_\star)\\
&\quad=\pr(\galaxymodel)\pr(\micromodel)\pr(\vltilde,\vbtilde|\galaxymodel)
\pr(\macromodel|\micromodel)
\\
&\qquad\times \pr(\va,\va_\star|\vltilde,\vbtilde,\vs,\vx_\star,\macromodel,\micromodel)
\pr(\vytilde,\vs,\vz|\vltilde,\vbtilde,\va,\micromodel,\galaxymodel)
,
\end{split}
\label{eqn:peverythingstar}
\end{equation}
which differs from~\eqref{eqn:peverything} only in the
$\pr(\va,\va_\star|\cdots)$ factor.
Marginalising,
\begin{equation}
  \begin{split}
 &
 \pr(\va_\star|\vytilde,\vltilde,\vbtilde,\vx_\star,\micromodel,\galaxymodel) 
\\
&\quad= \frac1{\pr(\vytilde,\vltilde,\vbtilde,\micromodel,\galaxymodel)}
\int\d\macromodel{\sum_{\vz}\int\d\vs\int\d\va\,
\pr(\vltilde,\vbtilde,\vs,\vytilde,\va,\va_\star,\vz,\macromodel,\micromodel,\galaxymodel|\vx_\star)},
  \end{split}
\label{eqn:map_dist}
\end{equation}
where, as in~\eqref{eqn:posteriormacro} above, much of the marginalisation can be carried out ``by hand''.

\subsection{Comparison with standard GP regression}
Notice that the pdf~\eqref{eqn:peverything} is a Gaussian in~$\va$: if the stellar distances~$\vs$ and membership  indicators~$\vz$ were known, then our extinction-mapping job would reduce to a straightforward GP regression problem.
In practice, however, stellar distances~$s$ are typically very uncertain.
One way of dealing with this is to project the distance uncertainties onto extinction by multiplying the distance uncertainties by the mean gradient of extinction with respect to distance, i.e., by using a first-order Taylor expansion.
This approach has been used by \cite{Vergely_FreireFerrero.2001} in their extinction maps.
\cite{Mchutchon_Rasmussen.2011} discuss, in significantly greater detail, a similar approach for the general GP regression problem.
This first-order Taylor expansion is a good approximation only when the uncertainties on distance are small.
In practice this is not the case, especially when the data used to estimate stellar distances is limited to photometry \citep{Green_Schlafly.2014, Sale_Magorrian.2015}.
Moreover, the mean gradient of extinction with respect to distance is unknown.
Therefore, we must instead sample the unknown distances of the stars in the learning phase.
In Paper~I this was achieved through the use of MCMC.
A similar approach was subsequently discussed by \cite{Cervone_Pillai.2015}.

As in standard GP regression, we do not know the precise (log) extinction~$a$ to each star, but instead possess only some noisy estimate of it.
This is much easier to deal with:
as shown in the derivation of equations \eqref{eqn:cond_GP} and~\eqref{eqn:cond_GP0} above, it is straightforward to account for measurement errors in the values of a Gaussian Process assuming the uncertainties are normally distributed.

These observational uncertainties are therefore straightforward to deal with, at least in principle.
The main {\it practical} difficulty in applying GP regression to the extinction mapping problem is sheer computational expense.
We are forced to infer or sample from the joint distributions that include $\vs$ and $\vz$.  Each such sample requires the  calculation of 
$\pr(\va|\vltilde,\vbtilde,\vs,\macromodel,\micromodel)$.   For a sample of $N$ stars this involves finding the inverse of an $N\times N$ matrix, which scales in CPU time as $\mathcal{O}(N^3)$ and in memory as $\mathcal{O}(N^2)$.
Given that we now live in an era of large surveys that contain $\sim 10^8 - 10^9$ objects, this is a clear barrier to applying the method of Paper~I.

\section{Accelerated extinction mapping}\label{sec:scheme}

We have just seen that the fundamental bottleneck in the scheme of Paper~I is the calculation of the joint PDF $\pr(\va|\vltilde,\vbtilde,\vs,\macromodel,\micromodel)$ of the extinctions to all $N$ stars in the catalogue.
Therefore it makes sense to look for ways of approximating this PDF.
Given the large size of modern survey catalogues, any such approximation should not only deliver significant speed and memory savings, but also allow easy parallelisation to allow the catalogue to be spread across multiple nodes.
In this section we describe a combination of two such schemes: Expectation Propagation \citep[EP][]{Minka_only.2001} and the Partially Independent Conditional approximation \citep[PIC][]{Snelson_Ghahramani.2007}.  Appendix~\ref{sec:accel} gives a more detailed overview of these and similar approaches for accelerating GP regression.

In order to simplify the exposition of our approach, we initially assume that the distances $\vs$ to the stars, their Gaussian mixture components $\vz$ and the dust microphysics $\micromodel$ are known exactly a priori.

\subsection{Acceleration using PIC and EP: simplified case}\label{sec:simplified}

Let us begin by focusing on the learning phase of GP regression.  In the simplified case in which we know $\vs$ and $\vz$ this reduces to estimating the posterior distribution of $\macromodel$, the hyperparameters that determine the large scale distribution of dust.
Employing the assumptions above allows us to simplify the posterior \eqref{eqn:peverything}, so that
\begin{equation}
\pr(\va, \macromodel | \vytilde) \propto \pr(\vytilde|\va) \pr(\va | \macromodel) \pr(\macromodel),
\end{equation}
and, marginalising,
\begin{equation}
\pr( \macromodel | \vytilde) \propto \int \, d\va\, \pr(\vytilde|\va) \pr(\va | \macromodel) \pr(\macromodel).
\end{equation}
For clarity we have suppressed the implicit conditioning on the assumed $(\vltilde,\vbtilde, \vs, \vz, \micromodel, \galaxymodel)$.

We begin by splitting the $N$ stars into  $K$ sub-catalogues, so that
\begin{equation}
\va = \begin{pmatrix}
\va_1 \\
\va_2 \\
\vdots \\\va_K
\end{pmatrix} ,
\end{equation}
where the $k^{\rm th}$ sub-catalogue contains $N_k$ stars and 
\begin{equation}
N = \sum_{k=1}^K N_k .
\end{equation}
$\vytilde$ is also partitioned into a matching set of $K$ sub-catalogues.
Typically, the partitioning is based on the on-sky position of the stars, so that sub-catalogues cover non-overlapping regions of the sky.

As we employ the Gaussian mixture model approximation \eqref{eqn:gmm_approx} and assume we know $\vs$ and $\vz$, the likelihood for each star follows
\begin{equation}
\tilde{a}_n | a_n \sim \mathcal{N} (a_n, \sigma_n),
\end{equation}
where $\tilde{a}_n$ is the `observed' log-extinction to the $n^{\rm th}$ star, so that $\vatilde$ is a sufficient statistic for $\vytilde$ in this context.
Further assuming that observations of one star are independent of observations of all other stars, implies that
\begin{equation}
\pr(\vytilde|\va) = \prod_{k=1}^{K} \pr(\vatilde_k|\va_k) \label{eqn:like_factor}\\ 
\end{equation}

Unfortunately, we cannot directly partition the expression for $\pr(\va|\macromodel)$ -- the bottleneck in our method -- in this way.
Because extinctions to neighbouring points in space are inevitably correlated, we instead have that
\begin{equation}
\pr(\va|\macromodel) = \prod_{k=1}^{K} \pr(\va_k|\va_1, \ldots \va_{k-1}, \macromodel)  , \label{eqn:nonPITC_extinction}
\end{equation}
which means that it is impossible
to partition a spatially contiguous catalogue into discrete, disjoint, uncorrelated sub-catalogues.
We see that the extinctions in the $k^{\rm th}$ sub-catalogue depend on $\macromodel$ \emph{and} the extinctions in the preceding $k-1$ sub-catalogues.
Calculating $\pr(\va|\macromodel)$ in this factorised form is no faster than in its basic unfactorised form.

Instead we make the partially independent (training) conditional approximation \citep[PI(T)C;][see also Appendix~\ref{sec:sparse}]{Snelson_Ghahramani.2007},  which involves introducing $M$ appropriately chosen {\it inducing points}, which are spatially distributed throughout the full catalogue.  The joint PDF is approximated as 
\begin{equation}
\pr(\va | \vu, \macromodel) \approx \prod_{k=1}^{K} \pr(\va_k | \vu, \macromodel) , \label{eqn:PITC_extinction}
\end{equation}
where $\vu$ gives the values of the GP at the full set of inducing points.
By making this PI(T)C approximation, we are assuming that the extinctions in different sub-catalogues are independent when conditioned on $\vu$ and $\macromodel$.
With this factorisation we have a cheap means of approximating $\pr(\va | \vu, \macromodel)$.
The inclusion of $\vu$ induces correlations among the values of $\va$ in different sub-catalogues.
We discuss the choice of locations of these inducing points later.

Given the full catalogue of observed extinctions~$\vatilde$, the joint posterior distribution of $\vu$ and~$\macromodel$ is then
\begin{equation}
\begin{split}
\pr(\vu, \macromodel | \vatilde) &\appropto \int \, d\va \pr(\macromodel) \pr(\vu | \macromodel) \prod_{k=1}^{K} \left(\pr(\vatilde_k|\va_k) \pr(\va_k | \vu, \macromodel) \right) \\
&= \pr(\macromodel) \pr(\vu | \macromodel) \prod_{k=1}^{K} \left(\int \, d\va_k \pr(\vatilde_k|\va_k) \pr(\va_k | \vu, \macromodel) \right).
\label{eq:EPposterior2fit}
\end{split}
\end{equation}
We next use the Expectation Propagation (EP) algorithm \citep{Minka_only.2001} to construct an approximation of the form
\begin{equation}
q(\vu, \macromodel) = \pr(\macromodel) \pr(\vu | \macromodel) \prod_{k=1}^{K} q_k(\vu, \macromodel)
\label{eqn:EPapprox}
\end{equation}
to this posterior.
The resulting $q_k(\vu, \macromodel)$ factors are known as {\it site} distributions.
Their functional form is chosen to suit the problem at hand.  Here we assume that each $q_k$ is some multivariate Gaussian.

Below we set out the steps involved in our implementation of EP.  The reader interested in {\it why} it works should consult Appendix~\ref{sec:EP} and references therein.

The EP algorithm uses a couple of sets of distributions that are constructed from the posterior~\eqref{eq:EPposterior2fit}.
The first is the set of  {\it cavity} distributions (see also~\ref{eqn:cavity}) obtained by 
omitting the $k^{\rm th}$ site distribution from the approximated posterior~\eqref{eqn:EPapprox}: 
\begin{equation}
q_{\_ k}(\vu, \macromodel) = \frac{1}{Z_k} \pr(\macromodel) \pr(\vu | \macromodel) \prod_{l \neq k} q_l(\vu, \macromodel),
\end{equation}
in which $Z_k$ is a normalisation factor.
As our assumed $q_k(\vu, \macromodel)$ is Gaussian and the factor 
$\pr(\vu | \macromodel)$ is by definition Gaussian, then as long as our prior $\pr(\macromodel)$ is Gaussian these cavity distributions will themselves be Gaussian.
We further define the {\it reduced cavity} distributions
\begin{equation}
\begin{split}
q'_{\_ k}(\vu, \macromodel) &= \frac{q_{\_ k}(\vu, \macromodel)}{\pr(\macromodel) \pr(\vu | \macromodel)} \\
&= \frac{1}{Z_k} \prod_{l \neq k} q_l(\vu, \macromodel) .
\label{eqn:reduced_cav}
\end{split}
\end{equation}
The second is the set of {\it tilted} distributions (see also~\ref{eqn:tilted})
\begin{equation}
\begin{split}
q_{\backslash k}(\vu, \macromodel) &= q'_{\_ k}(\vu, \macromodel) \pr(\macromodel) \pr(\vatilde_k, \vu | \macromodel) . 
\label{eqn:tiltedmain}
\end{split}
\end{equation}
Again, these are Gaussian.

After we make an initial guess for the site distributions $\{q_k\}$,
the EP algorithm proceeds as follows:

\begin{enumerate}
\item construct the sets of cavity and tilted distributions, $q_{\_k}$ and $q_{\backslash k}$;
\item for each $k=1,\ldots,K$, construct an improved estimate, $q^{\rm new}_k$, of the site distribution $q_k$ by matching the first- and second-order moments of $q^{\rm new}_kq_{\_k}$ to those of the tilted distribution $q_{\backslash k}/Z_k$.
\item set $q_k=q^{\rm new}_k$ and repeat until converged.
\end{enumerate}

\noindent The CPU and memory costs of carrying out this algorithm are dominated by the $\pr(\vatilde_k,\vu|\macromodel)$ factor in the tilted distribution~\eqref{eqn:tiltedmain}.
The CPU time needed therefore scales as $\mathcal{O}(N_i + M)^3$ and the memory as $\mathcal{O}(N_i + M)^2$.

With some further manipulation we can improve on this scaling, however.
Factorise the expensive $\pr(\vatilde_k,\vu|\macromodel)$ factor into the product of Gaussians
$\pr(\vu|\vatilde_k,\macromodel)\pr(\vatilde_k|\macromodel)$ and 
introduce $q'_{\backslash k}(\vu|\macromodel)$, defined via
\begin{equation}
q'_{\backslash k}(\vu,\macromodel)=q'_{\backslash k}(\vu|\macromodel)\pr(\macromodel).
\end{equation}
Then the tilted distribution~\eqref{eqn:tiltedmain} becomes
\begin{equation}
q_{\backslash k}(\vu, \macromodel) = q_{\_ k}(\macromodel) q'_{\_ k}(\vu | \macromodel) \pr(\vu | \vatilde_k, \macromodel) \pr(\vatilde_k | \macromodel).
\end{equation}
\def\tiltednormalizer{F}
The product of Gaussians $q'_{\_ k}(\vu | \macromodel) \pr(\vu | \vatilde_k, \macromodel)$ appearing here can be written 
\citep[e.g.][]{Murphy_only.2007} as a single Gaussian $\overline{q}_k(\vu | \vatilde_k, \hyper)$ multiplied by a normalisation factor
$\tiltednormalizer_k(\vatilde_k, \hyper)$.  So we then have
\begin{equation}
q_{\backslash k}(\vu, \macromodel) = \tiltednormalizer_k(\vatilde_k, \macromodel) q_{\_ k}(\macromodel) \overline{q}_k(\vu | \vatilde_i, \macromodel) \pr(\vatilde_k | \macromodel) . \label{eqn:simple_tilted}
\end{equation}
We provide explicit expressions for $\overline{q}_k(\vu | \vatilde_k, \hyper)$ and $\tiltednormalizer_k(\vatilde_k, \hyper)$ in Appendix~\ref{sec:momentmatching}

The CPU and memory costs associated with sampling from the tilted distribution in this form approximately scale as $\mathcal{O}(N_k^3)$ and $\mathcal{O}(N_k^2)$ respectively, an improvement over the formulation above.
Moreover, all of the terms in this form have simple and intuitive meanings:
\begin{itemize}
\item The cavity distribution $q_{\_ k}(\macromodel)$ is, by design, Gaussian and provides the constraints offered by the prior and the other sub-catalogues on $\macromodel$.
\item $\overline{q}_k(\vu | \vatilde_k, \macromodel)$ is a Gaussian distribution on $\vu$ whose mean and covariance depend on $\vatilde_k$ and $\macromodel$.
It is the predicted distribution of $\vu$ given the observations in the $k^{\rm th}$ sub-catalogue, the information provided by other catalogues through the cavity distribution and the GP prior that depends on $\macromodel$.
\item $\tiltednormalizer_k(\vatilde_k, \macromodel)$ measures the agreement between the prediction for $\vu$ from by the cavity distribution and the prediction implied by $\vatilde_k$.
\item Finally, $\pr(\vatilde_k | \macromodel)$ is the simple GP likelihood obtained by considering the $k^{\rm th}$ sub-catalogue in isolation.
\end{itemize}

We can also illustrate the significance of the factors in the cavity distribution by considering an alternative approximation.
Assume that the extinctions in different blocks are independent, when conditioned on $\macromodel$.  That is, suppose that, instead of the PITC approximation ~\eqref{eqn:PITC_extinction}, we assume that
\begin{equation}
\pr(\va, | \macromodel) \approx \prod_{k=1}^{K} \pr(\va_k | \macromodel),
\end{equation}
which is equivalent to a block-diagonal approximation to the covariance matrix, with the off diagonal blocks set to zero.
It is also equivalent to the PITC approximation with an empty set of inducing points.
Working through a similar derivation to that above, we obtain the tilted distribution
\begin{equation}
q_{\backslash k}(\macromodel) = q_{\_ k}(\macromodel) \pr(\vatilde_k | \macromodel) .
\end{equation}
Both factors in this block-diagonal tilted distribution also appear in the PITC tilted distribution~\eqref{eqn:simple_tilted}.
Thus, the two additional factors in the PITC tilted distribution~\eqref{eqn:simple_tilted} provide the approximation to the effect of the off-diagonal blocks in the covariance matrix.

Constructing and sampling from the tilted distribution is the most difficult stage in the application of EP to our simplified extinction mapping.
As all the prior distributions and site approximations are Gaussian, multiplying them to obtain the overall joint approximation $q(\vu, \macromodel)$ and subsequently finding the cavity distributions are both straightforward.
In addition, updating the site approximations can be done by simply moment matching to the tilted distributions.

\subsection{Acceleration using PIC and EP: general case}

We now consider how to carry out the learning phase of our GP extinction mapping in realistic cases, by relaxing some of the assumptions made in section~\ref{sec:simplified}.
In particular, we will make no assumptions about $\vs$ and $\vz$.
For the time being we will continue to assume a dust power spectrum $(\micromodel)$, whilst the assumption that the observed on-sky position of the stars is exactly correct is essentially always reasonable.

The fundamental approach we employ remains unchanged: we use a combination of the PIC approximation and EP.
Given that we no longer know the distances to the stars, $\vatilde$ is not a sufficient statistic for $\vytilde$.
Instead, given some $\vs$ and $\vz$, $(\vatilde, \vstilde)$ is a sufficient statistic, where $\log \tilde{s}_n$ is the mean log-distance of the component in the Gaussian mixture model approximation to the likelihood of the $n^{\rm th}$ star indicated by $\vz_n$, whilst $\tilde{a}_n$ is the mean log-extinction conditioned on $z_n$ and $s_n$.

We may split $\pr(\tilde y,s,z|a,\tilde l,\tilde b,\micromodel,\galaxymodel)$ into two factors,
\begin{equation}
\pr(\tilde y,s,z|a,\tilde l,\tilde b,\micromodel,\galaxymodel) = \pr(\tilde s, s, z|\tilde l,\tilde b,\galaxymodel) \pr(\tilde a,s,z|a,\tilde l,\tilde b,\micromodel,\galaxymodel).
\end{equation}
As we approximate the left hand side of this equation with a Gaussian mixture model~\eqref{eqn:gmm_approx}, as described in \cite{Sale_Magorrian.2015}, it follows that the right hand side can be approximated as the product of two Gaussians, where
\begin{align}
\log \tilde{s}_n(z_n) \sim \mathcal{N} (\log s_n, \sigma^{(s)}_n(z_n))\\
\tilde{a}_n(s_n, z_n)  \sim \mathcal{N} (a_n, \sigma^{(a)}_n(z_n)) .
\end{align}
Here $\tilde{s}_n(z_n)$ is the mean distance of the component in the GMM approximation to the likelihood indicated by $z_n$ and $\tilde{a}_n(s_n, z_n)$ the mean log-extinction implied by $z_n$ and $s_n$.

\subsubsection{Learning phase}

In addition to $\macromodel$ and $\vu$, we now also need to infer $\vs$ and $\vz$.
Fortunately, both $\vs$ and $\vz$ are `local' as described in section~\ref{sec:EP}: the observations in the $k^{\rm th}$ sub-catalogue depend only on $\vs_k$ and $\vz_k$ and are conditionally independent of $\vs_{j \neq k}$ and $\vz_{j \neq k}$.
Thus, the EP site, cavity and tilted approximations do not depend on $\vs$ or $\vz$.

By a similar derivation to that given in section~\ref{sec:simplified}, the tilted distributions are
\begin{equation}
\begin{split}
q_{\backslash k}(\vu, \macromodel) = \sum_{\vz_k}  \int & \, d\vs_k  q_{\_ k}(\macromodel) \overline{q}_k(\vu | \vatilde_k, \vs_k, \vz_k, \macromodel) \tiltednormalizer_k(\vatilde_k, \vs_k, \vz_k, \macromodel)  \\
& \pr(\vatilde_k | \macromodel, \vltilde, \vbtilde, \micromodel, \galaxymodel) \pr(\vstilde_k, \vs_k, \vz_k | \vltilde, \vbtilde, \galaxymodel).
\end{split} \label{eqn:full_tilted}
\end{equation}
This is the generalisation of  the tilted distribution in the simplified case~\eqref{eqn:simple_tilted}.
As we do not know $\vs_k$ nor $\vz_k$ a priori we have an additional factor that gives their probability, which we then marginalise.
We currently employ a Metropolis within Gibbs \citep{Tierney_only.1994} MCMC algorithm to sample from the intergrand above.
The marginalisation to obtain samples from $q_{\backslash k}(\vu, \macromodel)$ is then trivial.
A new site approximation is then obtained by moment matching.

\subsubsection{Prediction phase: mapmaking}

Having an estimate of the joint posterior distribution of $(\vs, \vz, \vu, \macromodel)$, we can embark on the prediction phase of GP regression and infer the extinction to a regular grid of points and thereby produce a 3d map of extinction.
In Appendix~\ref{app:prediction} we derive an approximation to the predictive distribution,
\begin{equation}
\begin{split}
& \pr(\va_{\star,k} | \vytilde, \vx_\star, \vltilde,\vbtilde,\micromodel,\galaxymodel) \\
&\quad\approx \iint \d\vu \, \d\macromodel \, \sum_{\vz_k} \int \d\vs_k \,  \Big( q_{\backslash k}(\vs_k, \vz_k,\vu,\macromodel) \\
&\qquad\qquad\qquad  \pr(\va_{\star,k}|\vs_k,\vz_k,\vu,\macromodel,\vatilde_k,\vltilde_k,\vbtilde_k,\micromodel,\galaxymodel) \Big),
\end{split}
\end{equation}
where $q_{\backslash k}(\vs_k, \vz_k,\vu,\macromodel)$ is the local unmarginalised tilted distribution for the $k^{\rm th}$ block of data as sampled from in each EP iteration.
Note, in particular, that we are now able to carry out the prediction phase, by focusing on each sub-catalogue in turn.
This allows trivial parallelisation of the training phase and is a direct consequence of combining the PIC approximation with EP.
The communication and smoothing between the different sub-catalogues is provided by the shared $\vu$ and $\macromodel$: the sub-catalogues are computationally independent while remaining statistically dependent.

In practice, we obtain samples from $q_{\backslash k}(\vs_k, \vz_k,\vu,\macromodel)$ using MCMC.
As a result, estimating the moments of $\pr(\va_{\star,k} | \vytilde, \vx_\star, \vltilde,\vbtilde,\micromodel,\galaxymodel)$ is easy.
Specifically, for each sampled $(\vs_k, \vz_k,\vu,\macromodel)$ we then obtain samples of $\va_{\star,k}$ from the Gaussian $\pr(\va_{\star,k}|\vs_k,\vz_k,\vu,\macromodel,\vatilde_k,\vltilde_k,\vbtilde_k,\micromodel,\galaxymodel)$.
We can then estimate the expectation of $\va_{\star,k}$ by taking the expectation of the samples of all the $\va_{\star,k}$ and similarly for higher moments.

\subsubsection{Scaling}

If we split a catalogue of $N$ stars into $K$ smaller catalogues, each of size $M$, then the total CPU cost of the training phase in the EP scheme will scale as $\mathcal{O}(N M^2)$, whilst memory costs will scale as $\mathcal{O}(NM)$.
As $M = N/K < N$, this is significantly more favourable than the $\mathcal{O}(N^3)$ and $\mathcal{O}(N^2)$ CPU and memory scaling of vanilla GP regression.
Meanwhile, the CPU costs of the prediction phase now scale as $\mathcal{O}(NM)$ rather than $\mathcal{O}(N^2)$.
In addition, both phases are trivially parallelisable, with minimal communication between processes only required when forming the new $q(\vu, \macromodel)$ at the end of each of the few (typically $\leq 5$) EP iterations.

In practice we can divide large observed catalogues into smaller sub-catalogues of roughly constant size, e.g. by having the sub-catalogues cover equal areas on the sky, 
Now if we add more similarly sized sub-regions to the map, the cost of producing larger extinction maps will scale linearly with the number of stars~$N$.
When allied to the fact that EP permits trivial parallelisation, with minimal communication between nodes, we now have an extinction mapping method that can be applied to larger catalogues, such as those that surveys such as Gaia are producing.
This vastly accelerated method relies on only two approximations: the PIC approximation and, in using EP, we have approximated the site distributions with a multivariate Gaussian $q_k(\macromodel, \vu)$.

The astute reader will have noticed that with very large catalogues we are essentially presented with two options: increase the number of inducing points or reduce the ratio of inducing points to observations.
If the spatial coverage of our data set is growing the second of these two options would result in the inducing points not being able to maintain an adequate spatial sampling relative to the scale of the covariance function.
As a consequence, there would be an increased loss of precision in the extinction maps produced and fingers of God would begin to appear as the correlation between blocks becomes weaker \citep{Vanhatalo_Pietilainen.2010}.
Therefore, we seek to avoid it as far as possible.
Increasing the number of inducing points would normally result in increased computation cost, just as increasing the number of observations increases the cost of vanilla GP regression.
Specifically, the contribution to the CPU time needed to invert the covariance matrix at of the $R$ inducing points, as needed in PIC based GP regression, scales as $\mathcal{O}(R^3)$.
When $R$ is small, this contribution is not significant relative to the $\mathcal{O}(NM^2)$ cost related to dealing with the observations.
However, given the cubic scaling, it can quickly become important as the number of inducing points increases.
Thus it would appear to be prohibitive to employ a large number of inducing points.

However, we note that it is possible to divide the total set of inducing points into a smaller number of subsets, $\vu = \{\vu_1 \ldots \vu_L\}$.
We can then apply the PIC approximation to the inducing points, employing a set of `hyper-inducing points' $\vv$.
In doing so we establish a clear hierarchy.
At the base we have our observations, divided into sub-catalogues.
A number of these sub-catalogues are tied together by a subset of inducing points under the PIC approximation.
Finally, all the subsets are related by again applying the PIC approximation and the hyper-inducing points.
This describes a three level hierarchy, though there is no reason it could not be extended to contain more levels as required by the size of the data.

A hierarchical PIC scheme of this form is well matched to computational infrastructure: one can easily imagine computational cluster nodes being hierarchically grouped in a similar manner.
Moreover, with HEALPix \citep{Gorski_Hivon.2005} it is easy to group observations into the required hierarchical structure.
Consequently, with a hierarchical PIC-EP scheme it would be possible to map extinction employing catalogues of essentially limitless size, with the CPU time and memory needed scaling only linearly with the total size of the catalogue and with the option of straightforward parallelisation.

\section{Tests with simulated Gaia data}\label{sec:verify}

\begin{figure*}
\includegraphics{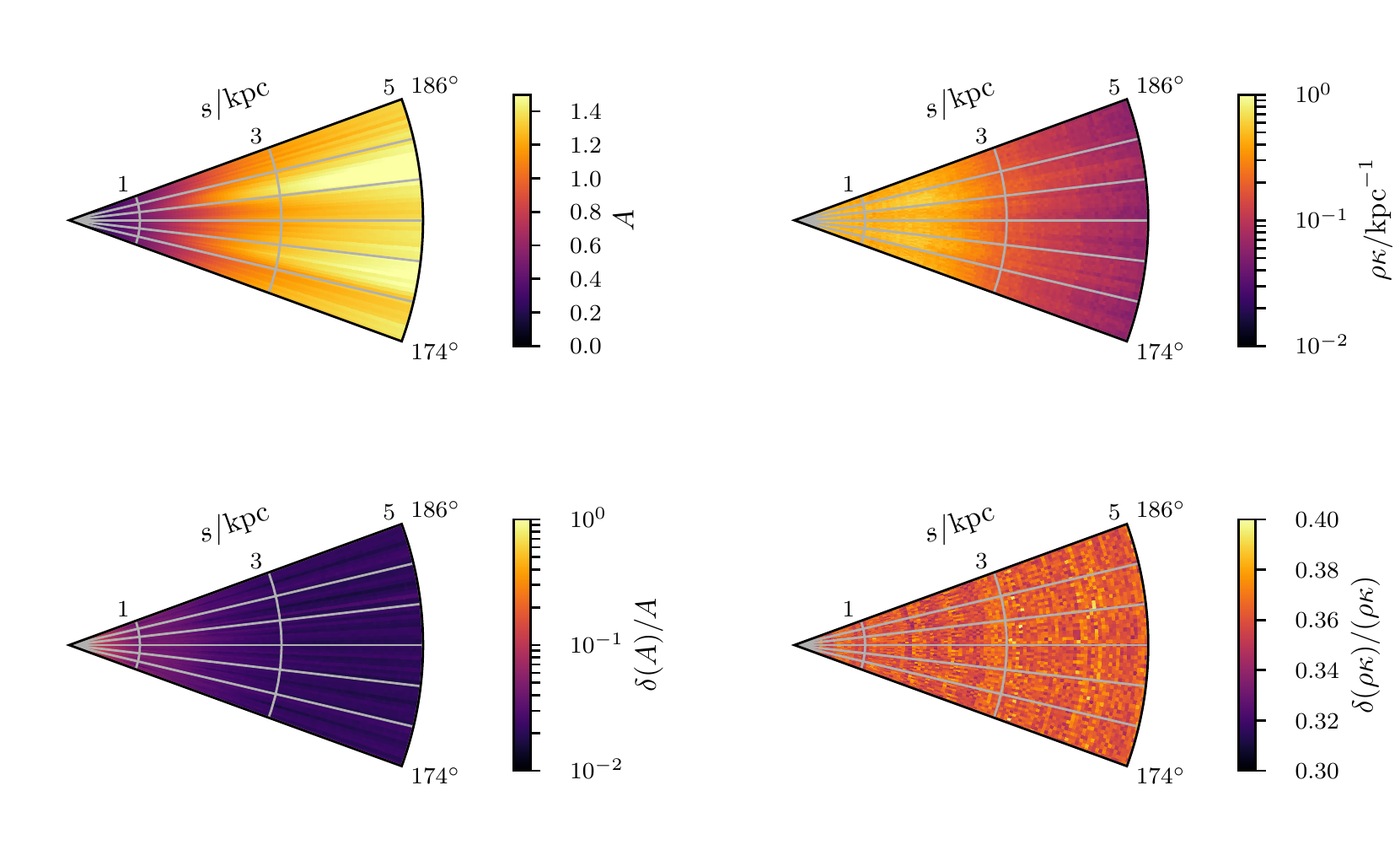}
\caption{An extinction map of the Galactic anticentre based on GOG \protect\citep{Luri_Palmer.2014} simulated data. The top left panel shows the estimated extinction, the top right the estimated pseudo-density, whilst the lower two panels show relative uncertainties.
  \label{fig:A_map}}
\end{figure*}

\begin{figure}
\includegraphics{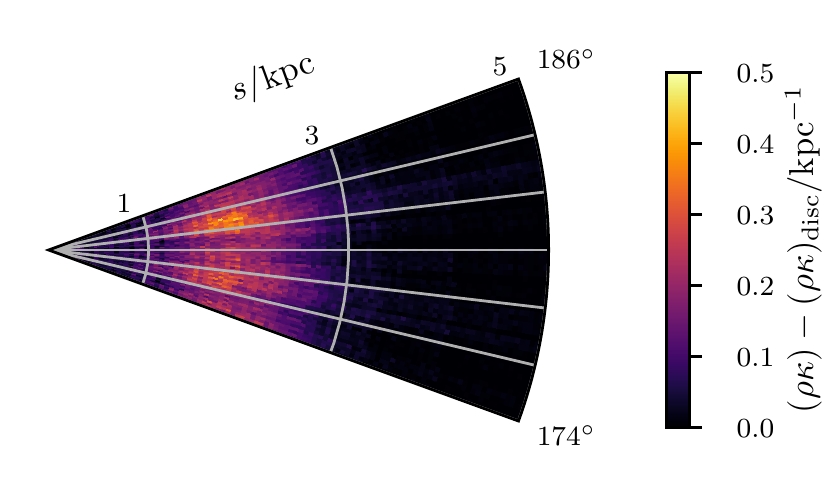}
\caption{The map of the estimated extinction pseudo-densities with the inferred contribution of the exponential disc of dust having been subtracted. The remaining extinction pseudo-density is due to the spiral arm.
  \label{fig:spiral_sub}}
\end{figure}

\begin{figure}
\includegraphics{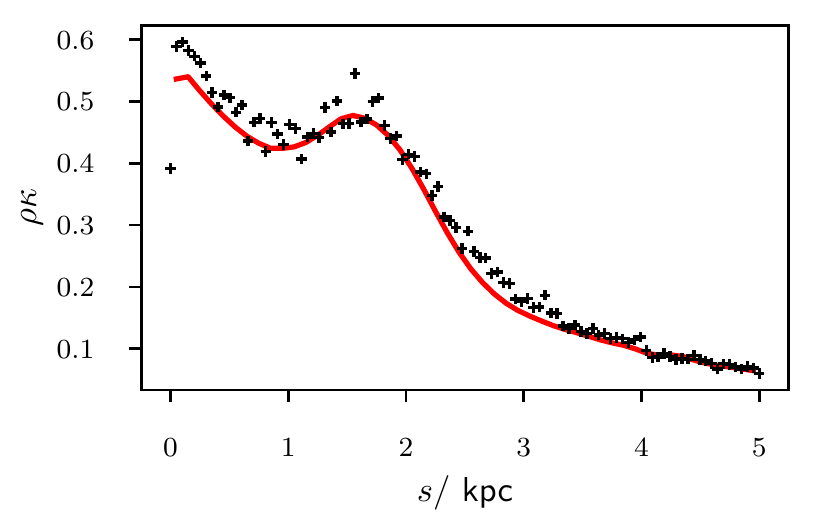}
\caption{The estimated extinction pseudo-densities along the sightline towards the anti-centre. The black crosses show the posterior estimates obtained, whilst the red line shows the pseudo-density from the D03 model.
  \label{fig:density}}
\end{figure}

\begin{figure*}
\includegraphics{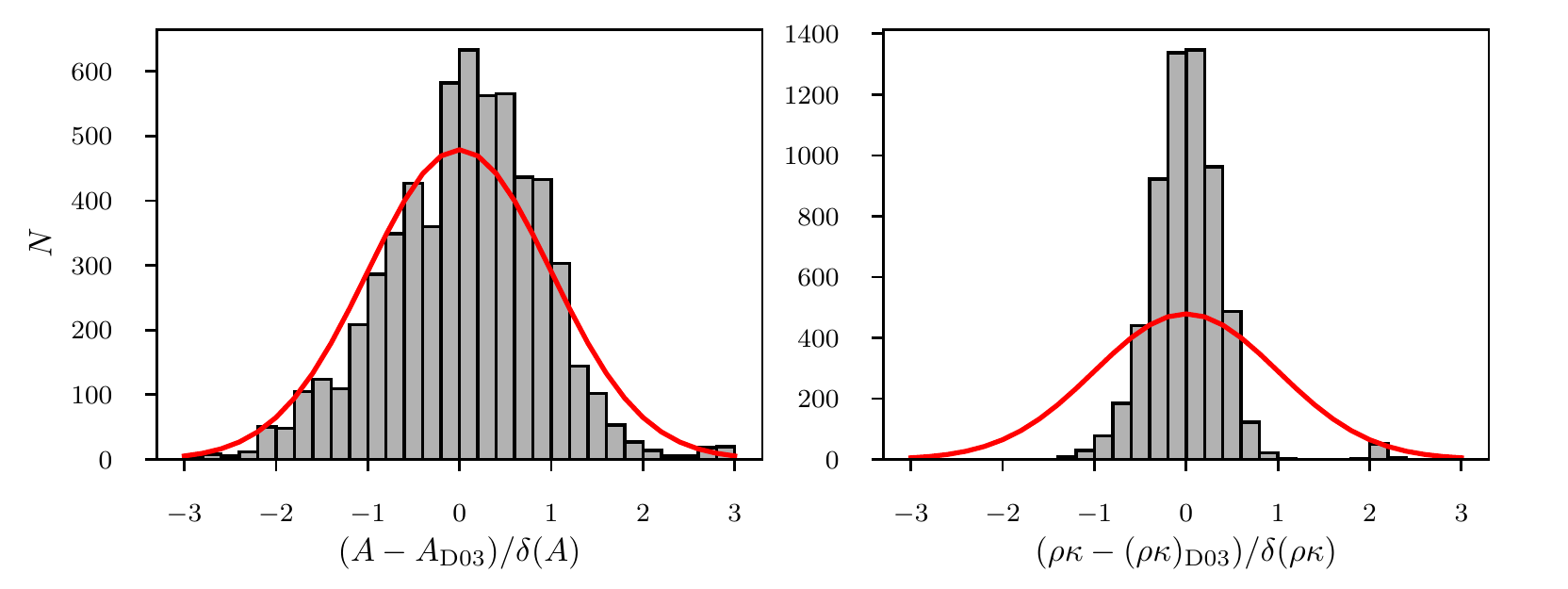}
\caption{A comparison between the estimated extinction map and the D03 model that is employed in the production of the simulated data. Each panel shows a histogram of the residuals divided by the measured uncertainties. The left panel compares extinctions and the right pseudo-densities. On each panel the standard Normal distribution has been overplotted with a red line.
  \label{fig:map_comp}}
\end{figure*}

\begin{figure}
\includegraphics{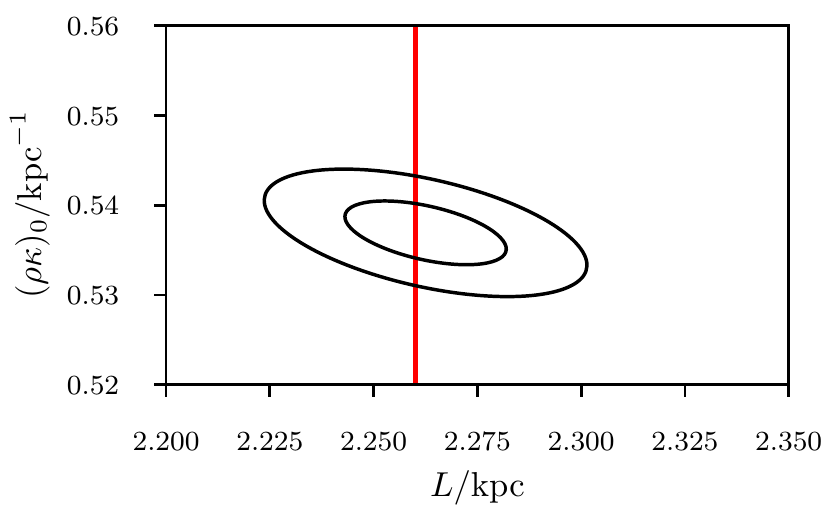}
\caption{The estimated posterior distribution on the disc scale length $L$ and the extinction pseudo-density at the sun. The red line shows the scale length employed in the D03 model.
  \label{fig:hyper_post_comp}}
\end{figure}

As in Paper~I, we employ simulated data to asses the quality of the inferences made our EP based algorithm.
Clearly Gaia will provide the most significant data data set for extinction mapping in the near future.
With that in mind, we test our implementation, which we name {\tt G-MEAD} (GP Mapping of Extinction Against Distance), on synthetic Gaia observations produced by the Gaia Object Generator \citep[GOG][]{Luri_Palmer.2014}.
GOG produces synthetic observations, including parallaxes and extinctions, for stars in the Gaia Universe Model \citep[GUM,][]{Robin_Luri.2012}.

The GUM employs the \citet[][hereafter D03]{Drimmel_Cabrera-Lavers.2003} Galactic dust model to describe extinction within the Galaxy.
The D03 model includes features such a spiral arms and normalises extinction along each sightline to that given by \cite{Schlegel_Finkbeiner.1998}.
It does not, however, include a direct simulation of interstellar turbulence.
As a result, it does not necessarily square with the GP based statistical model we employ and is not the perfect simulation on which to test our method.
However, we still consider the GOG simulations a useful test since they are otherwise a very faithful reproduction of what we might expect from the full Gaia catalogue.

In practical terms the non-simulation of interstellar turbulence has a number of effects.
First, we impose a relatively small value of the ratio between the standard deviation and mean of dust (3d) density.
If we set this value to zero we would be implicitly assuming that the distribution of dust (and so extinction) could be entirely described by our chosen mean function.
By setting it to a small value we are allowing a little variation from the mean.
This is necessary because the normalisation of the D03 map to \cite{Schlegel_Finkbeiner.1998} imposes some turbulent-like features on the data.
An effect of the small value of the variance ratio is that our model is less expressive than it otherwise might be.
This means it has reduced ability to describe features that do not appear in our mean function.
In addition, our statistical model is more complicated than that behind the data we wish to fit, thus we are at risk of over-fitting.
To mitigate this, we impose a floor on the covariance arrays of the cavity distributions.
 
Currently, our implementation contains no treatment for the (typically apparent-magnitude based) incompleteness of catalogues.
As demonstrated by \cite{Sale_only.2015} selection effects stemming from this incompleteness can have a severe and pathological impact on extinction maps.
Unfortunately, the approach developed in \cite{Sale_only.2015} cannot be directly applied here.
Therefore, we sidestep this issue by compiling a volume-limited test catalogue.
We achieve this by including only stars with masses greater than $1.2 M_{\odot}$ and distances less than 5000~pc.
In addition, we also discard the least informative stars, jettisoning those with relative parallax errors above one half.

Our test catalogue ultimately contains 6349 stars within a one square degree are defined by $ 174^{\circ} \leq l < 186 ^{\circ}$ and $ |b| < 2.5\arcmin$.
We adopt this `letter-box' area to make it easier to plot maps on paper.
We then sub-divide this into 36 sub-catalogues of $ 20 \arcmin \times 5 \arcmin$.
A total of 140 inducing points are placed along the 35 catalogue boundaries at distances of $1, 2, 3,$ and $4$~kpc.
By placing the inducing points along the sub-catalogue boundaries we minimise the discontinuities that could otherwise occur there \citep{Vanhatalo_Pietilainen.2010}.

In addition to mapping the extinction $A$, we would also like to infer the 3d distribution of the dust density.
However, as 
\begin{equation}
\frac{\d A}{\d s} = \rho \kappa ,
\end{equation}
where $\rho$ is the dust density and $\kappa$ its opacity, it is difficult to extract $\rho$ directly.
The dust opacity varies with changes in the dust grain size distribution and with dust composition, which means it is not completely straightforward to unpick its influence.
So, we instead we map the extinction pseudo-density $\rho \kappa$, which can be easily determined.

We adopt a mean function that assumes dust is distributed as an exponential disc to which we add a Gaussian `bump' that roughly approximates a spiral arm.
As a result, there are five hyperparameters.
Three describe the disc: the dust scale length $L$, scale height $H$\footnote{We note that by choosing a region in the mid-plane the dust scale height is almost completely unconstrained.} and the pseudo-density of extinction at the sun $(\rho \kappa)_0$.
A further two describe the bump: its distance $s_{\rm bump}$ and total integrated extinction $A_{\rm bump}$; we assume the width of the bump to be fixed with a 400~pc standard deviation.
We apply a log-normal hyperprior to each of these 5 parameters with means of $3000$~pc for $L$, $150$~pc for $H$, $0.75$~kpc$^{-1}$ for $(\rho \kappa)_0$, $2$~kpc for $s_{\rm bump}$ and 0.1 for $A_{\rm bump}$.
The standard deviation of each hyperprior is set equal to its mean value, so that the hyperprior is largely uninformative.

Fig.~\ref{fig:A_map} shows the result obtained by running {\tt G-MEAD} on this catalogue.
The estimated uncertainties on the extinction map are very small, typically on the order of a few percent.
This is more than an order of magnitude smaller than comparable uncertainties in existing maps such as \cite{Sale_Drew.2014} and \cite{Green_Schlafly.2015}.
To an extent, this is a result of the simulated data not including a proper description of interstellar turbulence, although we still consider this to be an example of the precisions that may be achievable with Gaia data.

The map successfully captures the the spiral arm that is present in the D03 model.
Its relatively weak strength makes it difficult to pick out in Fig.~\ref{fig:A_map}, but when the contribution of the exponential disc is subtracted or when looking at individual sightlines, as in Figs.~\ref{fig:spiral_sub} and~\ref{fig:density}, it is much more obvious.

There are no significant `fingers of God' in the plot of pseudo-density and extinction varies smoothly.
However, low-level deviations between sightlines can be seen in the pseudo-density plot.
But, we note that the strength of these variations are well below the estimated noise.
These slight deviations appear because the density map within the area spanned by a given sub-catalogue is only directly conditioned on the observations of that sub-catalogue;
the noise characteristics vary somewhat between sub-catalogues (e.g. as the number of stars and their distance distribution changes) resulting in changes in to the width of the typically skewed pdf of pseudo-density and so in the posterior expectations of pseudo-density.
These small differences are then enhanced slightly further by sampling noise.

We also make a comparison, in Fig.~\ref{fig:map_comp}, to the D03 Galactic extinction model, on which the simulations are based, showing an excellent agreement.
It is also apparent that, whilst the extinction uncertainties appear reasonable, the uncertainties on density have been somewhat over-estimated and are actually on the order of $10\%$.
This is an extraordinarily high level of precision, but we again caution that with real data uncertainties will likely be somewhat higher.

In addition, we can estimate the posterior distributions of the hyperparameters that define the exponential dust disc.
In Fig.~\ref{fig:hyper_post_comp} we plot this distribution for the dust scale length and normalisation density.
The posterior distribution of the scale length is in near exact agreement with the value used in the D03 model.

For completeness, we must clarify exactly what Fig.~\ref{fig:A_map} and similar are showing.
For extinction we plot maps of the mean and variance of the \textit{marginal} posterior predictive $\pr(a_{\star} | \vx_{\star}, \vytilde, \vltilde, \vbtilde, \micromodel)$ on a regular grid of points.
Note that we do not impose the condition that extinction must increase with distance. 
Estimating the map of density involves finding the joint posterior predictive at two locations for each grid point: the grid point itself and another point in the same direction but 10~pc further away.
Then, the density is simply found by sampling from the difference between the two extinctions and imposing the condition that the extinction to the further point is greater than that to the nearer.

Roughly 2 hours of time on an Amazon AWS t2.micro instance, with 1 (virtual) CPU, was required to produce the map shown in Fig.~\ref{fig:A_map}.
Given that the region covered contains approximately 1 in every $10^4$ stars in the GOG catalogue, we estimate that on the order of $10^4$ CPU hours would be required to map the entire Gaia catalogue.
This is well within the reach of reasonably sized CPU clusters.
By way of contrast, we estimate that roughly $10^{11}$ CPU years would be required to map a similar sized catalogue using vanilla GP regression, which is therefore, obviously, utterly unachievable.

Our input catalogue contained cuts on the basis of the `true' distances and masses of the stars, discarding nearly $90\%$ of sources.
With real Gaia data it will obviously not be possible to select the catalogue on this basis.
However, it is clearly the case that some stars convey more information about the 3d distribution of extinction than others.
Judicious use of e.g. photometric colour/magnitude cuts should enable us to preferentially select the most informative stars and so discard a large proportion of the Gaia catalogue.

\subsection{Conditioning on $dA/ds \geq 0$}

\begin{figure}
\includegraphics{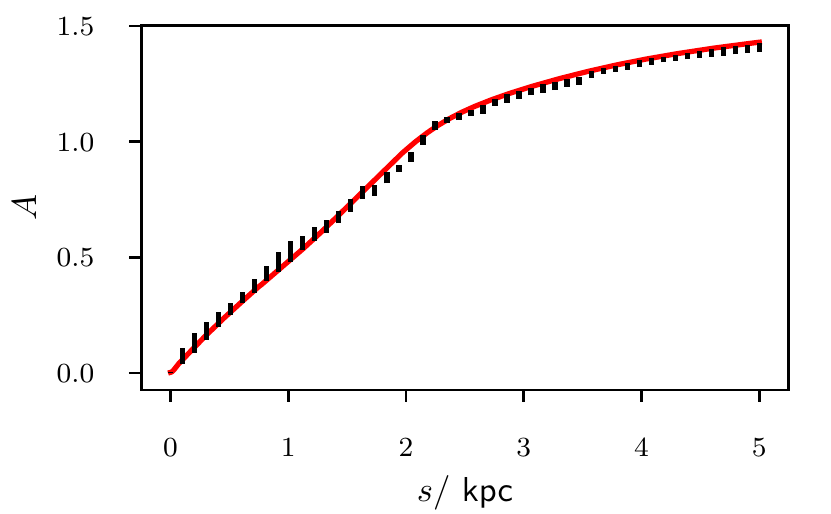}
\caption{Posterior estimates with uncertainties of the extinction along the sightline $(l, b) = (174.167^{\circ}, 0^{\circ})$ plotted as black error bars. Note that the small uncertainties make it difficult to resolve some of the error bars.
This is compared to the D03 model (red line) that is employed in the production of the simulated data
  \label{fig:condition_comp}}
\end{figure}

As discussed above, the maps in Fig.~\ref{fig:A_map} do not enforce the requirement that (monochromatic) extinction must increase with distance\footnote{Formally this requirement is not absolute, being breached in reflection nebulae for example.
However, this contribution typically only occurs over limited wavelength ranges and the pipeline processing of photometric data will often include the subtraction of diffuse emission from extended sources.
So, this requirement does hold in practice.}. 
Were we to condition our maps on $dA/ds \geq 0$, or equivalently $da/ds \geq 0$\footnote{Since we assume $A \geq 0$}, we would be providing extra information that would increase the precision of the maps.

We now consider a line of points along a sightline $\vx_{\star}$.
The posterior predictive distribution $\pr(\va_{\star} | \vx_{\star}, \vytilde, \vltilde, \vbtilde, \micromodel)$ for the extinction to these points is multivariate normal, following the approximation described in Appendix~\ref{app:prediction}.
The constraint that $da/ds \geq 0$ truncates this distribution, such that $\pr(\va_{\star} | \vx_{\star}, \vytilde, \vltilde, \vbtilde, \micromodel, da/ds \geq 0)$ follows a truncated multivariate normal distribution.

We use the method of \cite{Pakman_Paninski.2012} to sample from the truncated multivariate normal posterior predictive distribution.
\cite{Pakman_Paninski.2012} employ a Hamiltonian Monte Carlo approach, with reflection at the truncation boundaries that produces samples in orders of magnitude less time than more straightforward approaches, such as rejection sampling or simple MCMC.
Given a sample from the truncated multivariate normal posterior predictive distribution it is then trivial to calculate the moments.

Fig.~\ref{fig:condition_comp} demonstrates the effect of conditioning on $dA/ds \geq 0$ for the example in the previous section.
Each data point in this plot shows the mean and standard deviation of $\pr(a_{\star} | \vx_{\star}, \vytilde, \vltilde, \vbtilde, \micromodel, da/ds \geq 0)$, the posterior distribution of extinction conditioned on $dA/ds \geq 0$.
The effect of conditioning on $da/ds \geq 0$ is significant, typically providing a 2--fold improvement in the precision.
However, this improvement is limited by the fact that the uncertainty in the unconditioned map is relatively low, due to the lack of small scale turbulent structure in the simulated ISM.
With real data we would expect a larger improvement in the precision.

\section{Closing discussion}\label{sec:close}

In Paper~I we presented a method for mapping extinction in three dimensions using GPs.
This statistical description is an excellent match to the physics of the ISM: it reproduces the observed density and column density distributions, allows the incorporation of a model of interstellar turbulence and neatly delineates the split between the small scale physics of the ISM that is driven by turbulence and the larger scale physics.
Unfortunately, exact GP regression scales poorly to large datasets, with the CPU time needed scaling as $\mathcal{O}(N^3)$.
Consequently, processing a Gaia-sized catalogue with the unmodified algorithm of Paper~I would have required an unfeasibly long time.

In section~\ref{sec:scheme}, we describe a scheme by which we can dramatically accelerate the inference of GP-based extinction maps.
Our approach is based on the combination of two approximations: the PIC approximation \citep{Snelson_Ghahramani.2007} and Expectation Propagation \citep{Minka_only.2001}.
The principal benefit of employing these two approximations is that they dramatically decrease the CPU time and memory needed to infer the extinction map, with both becoming an approximating linear function of the number of stars studied.
In addition, the method significantly reduces the amount of network communication required if the construction of the map is distributed across multiple computing nodes and enables different portions of the map to be computed asynchronously.
Other solutions for accelerating GP regression are available, but the method we have described is fast, trivially parallelisable, and conceptually and computationally straightforward.

We have implemented our scheme in a library {\tt G-MEAD}. We have shown, using simulated Gaia data, that {\tt G-MEAD} can produce 3d extinction maps with hitherto unmatched precision and accuracy.
In addition, we can easily recognise features such as spiral arms that have eluded previous maps.
When applied to real Gaia data the resultant map will not only allow the direct study of the Galactic ISM, but will also support the wider study of the Galaxy, the stars that comprise it and the physical processes at work.

As we have described, it is now feasible to apply our method to a catalogue of a size similar to that which Gaia will produce.
However, selection effects have a pathological impact on extinction mapping \citep{Sale_only.2015} and we currently lack a method for dealing with them when mapping extinction with GPs.
As a result, the improvements needed to overcome this hurdle will be the subject of a future paper.
Once this has been overcome, the GP based method we have described will be able to produce 3d extinction maps from Gaia that will exhibit unparalleled detail, precision and accuracy.

\section*{Acknowledgements}

We thank James Binney and Ralph Sch\"onrich for comments on an earlier draft of this paper.
The research leading to the results presented here was supported by
the United Kingdom Science Technology and Facilities Council (STFC,
ST/K00106X/1, ST/M00127X/1), the European Research Council under the European
Union’s Seventh Framework Programme (FP7/2007-2013)/ERC grant
agreement no. 321067.

\bibliography{bibliography-2,ml,astroph_3,temp}

\newcommand{\noop}[1]{}
\begin{thebibliography}{}
\makeatletter
\relax
\def\mn@urlcharsother{\let\do\@makeother \do\$\do\&\do\#\do\^\do\_\do\%\do\~}
\def\mn@doi{\begingroup\mn@urlcharsother \@ifnextchar [ {\mn@doi@}
  {\mn@doi@[]}}
\def\mn@doi@[#1]#2{\def\@tempa{#1}\ifx\@tempa\@empty \href
  {http://dx.doi.org/#2} {doi:#2}\else \href {http://dx.doi.org/#2} {#1}\fi
  \endgroup}
\def\mn@eprint#1#2{\mn@eprint@#1:#2::\@nil}
\def\mn@eprint@arXiv#1{\href {http://arxiv.org/abs/#1} {{\tt arXiv:#1}}}
\def\mn@eprint@dblp#1{\href {http://dblp.uni-trier.de/rec/bibtex/#1.xml}
  {dblp:#1}}
\def\mn@eprint@#1:#2:#3:#4\@nil{\def\@tempa {#1}\def\@tempb {#2}\def\@tempc
  {#3}\ifx \@tempc \@empty \let \@tempc \@tempb \let \@tempb \@tempa \fi \ifx
  \@tempb \@empty \def\@tempb {arXiv}\fi \@ifundefined
  {mn@eprint@\@tempb}{\@tempb:\@tempc}{\expandafter \expandafter \csname
  mn@eprint@\@tempb\endcsname \expandafter{\@tempc}}}

\bibitem[\protect\citeauthoryear{Ambikasaran \& Darve}{Ambikasaran \&
  Darve}{2013}]{Ambikasaran_Darve.2013}
Ambikasaran S.,  Darve E.,  2013, Journal of Scientific Computing, pp 1--25

\bibitem[\protect\citeauthoryear{Ambikasaran, Foreman-Mackey, Greengard, Hogg
  \& O’Neil}{Ambikasaran et~al.}{2016}]{Ambikasaran_Foreman-Mackey.2016}
Ambikasaran S.,  Foreman-Mackey D.,  Greengard L.,  Hogg D.~W.,   O’Neil M.,
  2016, IEEE transactions on pattern analysis and machine intelligence, 38, 252

\bibitem[\protect\citeauthoryear{Bishop}{Bishop}{2006}]{Bishop_only.2006}
Bishop C.~M.,  2006, Pattern Recognition and Machine Learning (Information
  Science and Statistics).
Springer-Verlag New York, Inc., Secaucus, NJ, USA

\bibitem[\protect\citeauthoryear{{Cervone} \& {Pillai}}{{Cervone} \&
  {Pillai}}{2015}]{Cervone_Pillai.2015}
{Cervone} D.,  {Pillai} N.~S.,  2015, preprint, \href
  {http://adsabs.harvard.edu/abs/2015arXiv150608256C} {} (\mn@eprint {arXiv}
  {1506.08256})

\bibitem[\protect\citeauthoryear{{Chen}, {Schultheis}, {Jiang}, {Gonzalez},
  {Robin}, {Rejkuba}  \& {Minniti}}{{Chen} et~al.}{2013}]{Chen_Schultheis.2013}
{Chen} B.~Q.,  {Schultheis} M.,  {Jiang} B.~W.,  {Gonzalez} O.~A.,  {Robin}
  A.~C.,  {Rejkuba} M.,   {Minniti} D.,  2013, \mn@doi [\aap]
  {10.1051/0004-6361/201219682}, \href
  {http://adsabs.harvard.edu/abs/2013A%26A...550A..42C} {550, A42}

\bibitem[\protect\citeauthoryear{Cornford, Csato  \& Opper}{Cornford
  et~al.}{2005}]{Cornford_Csato.2005}
Cornford D.,  Csato L.,   Opper M.,  2005, \mn@doi [Geographical Analysis]
  {10.1111/j.1538-4632.2005.00635.x}, 37, 183

\bibitem[\protect\citeauthoryear{{Deisenroth} \& {Ng}}{{Deisenroth} \&
  {Ng}}{2015}]{Deisenroth_Ng.2015}
{Deisenroth} M.~P.,  {Ng} J.~W.,  2015, preprint, \href
  {http://adsabs.harvard.edu/abs/2015arXiv150202843D} {} (\mn@eprint {arXiv}
  {1502.02843})

\bibitem[\protect\citeauthoryear{{Drimmel}, {Cabrera-Lavers}  \&
  {L{\'o}pez-Corredoira}}{{Drimmel} et~al.}{2003}]{Drimmel_Cabrera-Lavers.2003}
{Drimmel} R.,  {Cabrera-Lavers} A.,   {L{\'o}pez-Corredoira} M.,  2003, \mn@doi
  [\aap] {10.1051/0004-6361:20031070}, \href
  {http://adsabs.harvard.edu/abs/2003A%26A...409..205D} {409, 205}

\bibitem[\protect\citeauthoryear{Frey \& Osborne}{Frey \&
  Osborne}{2013}]{Frey_Osborne.2013}
Frey C.~B.,  Osborne M.~A.,  2013, Technical report, The future of employment:
  how susceptible are jobs to computerisation?

\bibitem[\protect\citeauthoryear{{Gaia Collaboration} et~al.,}{{Gaia
  Collaboration} et~al.}{2016}]{Gaia_Prusti.2016}
{Gaia Collaboration} et~al., 2016, \mn@doi [\aap]
  {10.1051/0004-6361/201629272}, \href
  {http://adsabs.harvard.edu/abs/2016A%26A...595A...1G} {595, A1}

\bibitem[\protect\citeauthoryear{Gelman, Vehtari, Jyl{\"a}nki, Robert, Chopin
  \& Cunningham}{Gelman et~al.}{2014}]{Gelman_Vehtari.2015}
Gelman A.,  Vehtari A.,  Jyl{\"a}nki P.,  Robert C.,  Chopin N.,   Cunningham
  J.~P.,  2014, arXiv preprint arXiv:1412.4869

\bibitem[\protect\citeauthoryear{{G{\'o}rski}, {Hivon}, {Banday}, {Wandelt},
  {Hansen}, {Reinecke}  \& {Bartelmann}}{{G{\'o}rski}
  et~al.}{2005}]{Gorski_Hivon.2005}
{G{\'o}rski} K.~M.,  {Hivon} E.,  {Banday} A.~J.,  {Wandelt} B.~D.,  {Hansen}
  F.~K.,  {Reinecke} M.,   {Bartelmann} M.,  2005, \mn@doi [\apj]
  {10.1086/427976}, \href {http://adsabs.harvard.edu/abs/2005ApJ...622..759G}
  {622, 759}

\bibitem[\protect\citeauthoryear{{Green} et~al.,}{{Green}
  et~al.}{2014}]{Green_Schlafly.2014}
{Green} G.~M.,  et~al., 2014, \mn@doi [\apj] {10.1088/0004-637X/783/2/114},
  \href {http://adsabs.harvard.edu/abs/2014ApJ...783..114G} {783, 114}

\bibitem[\protect\citeauthoryear{{Green} et~al.,}{{Green}
  et~al.}{2015}]{Green_Schlafly.2015}
{Green} G.~M.,  et~al., 2015, \mn@doi [\apj] {10.1088/0004-637X/810/1/25},
  \href {http://adsabs.harvard.edu/abs/2015ApJ...810...25G} {810, 25}

\bibitem[\protect\citeauthoryear{Hensman, Fusi  \& Lawrence}{Hensman
  et~al.}{2013}]{Hensman_Fusi.2013}
Hensman J.,  Fusi N.,   Lawrence N.~D.,  2013, in Conference on Uncertainty in
  Artificial Intellegence. pp 282--290

\bibitem[\protect\citeauthoryear{Hinton}{Hinton}{2002}]{Hinton_only.2002}
Hinton G.~E.,  2002, Neural computation, 14, 1771

\bibitem[\protect\citeauthoryear{{Kolmogorov}}{{Kolmogorov}}{1941}]{Kolmogorov_only.1941}
{Kolmogorov} A.,  1941, Akademiia Nauk SSSR Doklady, \href
  {http://adsabs.harvard.edu/abs/1941DoSSR..30..301K} {30, 301}

\bibitem[\protect\citeauthoryear{Lindgren, Rue  \& Lindstr{\"o}m}{Lindgren
  et~al.}{2011}]{Lindgren_Rue.2011}
Lindgren F.,  Rue H.,   Lindstr{\"o}m J.,  2011, Journal of the Royal
  Statistical Society: Series B (Statistical Methodology), 73, 423

\bibitem[\protect\citeauthoryear{{Luri} et~al.,}{{Luri}
  et~al.}{2014}]{Luri_Palmer.2014}
{Luri} X.,  et~al., 2014, \mn@doi [\aap] {10.1051/0004-6361/201423636}, \href
  {http://adsabs.harvard.edu/abs/2014A%26A...566A.119L} {566, A119}

\bibitem[\protect\citeauthoryear{{Marshall}, {Robin}, {Reyl{\'e}}, {Schultheis}
   \& {Picaud}}{{Marshall} et~al.}{2006}]{Marshall_Robin.2006}
{Marshall} D.~J.,  {Robin} A.~C.,  {Reyl{\'e}} C.,  {Schultheis} M.,   {Picaud}
  S.,  2006, \mn@doi [\aap] {10.1051/0004-6361:20053842}, \href
  {http://adsabs.harvard.edu/abs/2006A%26A...453..635M} {453, 635}

\bibitem[\protect\citeauthoryear{McHutchon \& Rasmussen}{McHutchon \&
  Rasmussen}{2011}]{Mchutchon_Rasmussen.2011}
McHutchon A.,  Rasmussen C.~E.,  2011, in Advances in Neural Information
  Processing Systems. pp 1341--1349

\bibitem[\protect\citeauthoryear{Minka}{Minka}{2001}]{Minka_only.2001}
Minka T.~P.,  2001, in Proceedings of the Seventeenth conference on Uncertainty
  in artificial intelligence. pp 362--369

\bibitem[\protect\citeauthoryear{Murphy}{Murphy}{2007}]{Murphy_only.2007}
Murphy K.~P.,  2007, Technical report, Conjugate Bayesian analysis of the
  Gaussian distribution.
Department of Computer Science, University of British Columbia

\bibitem[\protect\citeauthoryear{{Nordlund} \& {Padoan}}{{Nordlund} \&
  {Padoan}}{1999}]{Nordlund_Padoan.1999}
{Nordlund} {\AA}.~K.,  {Padoan} P.,  1999, in {Franco} J.,  {Carraminana} A.,
  eds, Interstellar Turbulence. p.~218 (\mn@eprint {} {astro-ph/9810074})

\bibitem[\protect\citeauthoryear{{Ostriker}, {Stone}  \& {Gammie}}{{Ostriker}
  et~al.}{2001}]{Ostriker_Stone.2001}
{Ostriker} E.~C.,  {Stone} J.~M.,   {Gammie} C.~F.,  2001, \mn@doi [\apj]
  {10.1086/318290}, \href {http://adsabs.harvard.edu/abs/2001ApJ...546..980O}
  {546, 980}

\bibitem[\protect\citeauthoryear{Pakman \& Paninski}{Pakman \&
  Paninski}{2014}]{Pakman_Paninski.2012}
Pakman A.,  Paninski L.,  2014, \mn@doi [Journal of Computational and Graphical
  Statistics] {10.1080/10618600.2013.788448}, 23, 518

\bibitem[\protect\citeauthoryear{Parts, Stegle, Winn  \& Durbin}{Parts
  et~al.}{2011}]{Parts_Stegle.2011}
Parts L.,  Stegle O.,  Winn J.,   Durbin R.,  2011, \mn@doi [PLoS Genet]
  {10.1371/journal.pgen.1001276}, 7, e1001276

\bibitem[\protect\citeauthoryear{Qui{\~n}onero-Candela \&
  Rasmussen}{Qui{\~n}onero-Candela \&
  Rasmussen}{2005}]{Quinonero-Candela_Rasmussen.2005}
Qui{\~n}onero-Candela J.,  Rasmussen C.~E.,  2005, The Journal of Machine
  Learning Research, 6, 1939

\bibitem[\protect\citeauthoryear{Rasmussen \& Williams}{Rasmussen \&
  Williams}{2005}]{Rasmussen_Williams.2005}
Rasmussen C.~E.,  Williams C. K.~I.,  2005, Gaussian Processes for Machine
  Learning (Adaptive Computation and Machine Learning).
The MIT Press

\bibitem[\protect\citeauthoryear{{Rezaei Kh.}, {Bailer-Jones}, {Hanson}  \&
  {Fouesneau}}{{Rezaei Kh.} et~al.}{2017}]{RezaeiKh._Bailer-Jones.2017}
{Rezaei Kh.} S.,  {Bailer-Jones} C.~A.~L.,  {Hanson} R.~J.,   {Fouesneau} M.,
  2017, \mn@doi [\aap] {10.1051/0004-6361/201628885}, \href
  {http://adsabs.harvard.edu/abs/2017A%26A...598A.125R} {598, A125}

\bibitem[\protect\citeauthoryear{{Robin} et~al.,}{{Robin}
  et~al.}{2012}]{Robin_Luri.2012}
{Robin} A.~C.,  et~al., 2012, \mn@doi [\aap] {10.1051/0004-6361/201118646},
  \href {http://adsabs.harvard.edu/abs/2012A%26A...543A.100R} {543, A100}

\bibitem[\protect\citeauthoryear{Rue, Martino  \& Chopin}{Rue
  et~al.}{2009}]{Rue_Martino.2009}
Rue H.,  Martino S.,   Chopin N.,  2009, Journal of the royal statistical
  society: Series b (statistical methodology), 71, 319

\bibitem[\protect\citeauthoryear{{Sale}}{{Sale}}{2015}]{Sale_only.2015}
{Sale} S.~E.,  2015, \mn@doi [\mnras] {10.1093/mnras/stv1459}, \href
  {http://adsabs.harvard.edu/abs/2015MNRAS.452.2960S} {452, 2960}

\bibitem[\protect\citeauthoryear{{Sale} \& {Magorrian}}{{Sale} \&
  {Magorrian}}{2014}]{Sale_Magorrian.2014}
{Sale} S.~E.,  {Magorrian} J.,  2014, \mn@doi [\mnras] {10.1093/mnras/stu1728},
  \href {http://adsabs.harvard.edu/abs/2014MNRAS.445..256S} {445, 256}

\bibitem[\protect\citeauthoryear{{Sale} \& {Magorrian}}{{Sale} \&
  {Magorrian}}{2015}]{Sale_Magorrian.2015}
{Sale} S.~E.,  {Magorrian} J.,  2015, \mn@doi [\mnras] {10.1093/mnras/stv068},
  \href {http://adsabs.harvard.edu/abs/2015MNRAS.448.1738S} {448, 1738}

\bibitem[\protect\citeauthoryear{{Sale} et~al.,}{{Sale}
  et~al.}{2014}]{Sale_Drew.2014}
{Sale} S.~E.,  et~al., 2014, \mn@doi [\mnras] {10.1093/mnras/stu1090}, \href
  {http://adsabs.harvard.edu/abs/2014MNRAS.443.2907S} {443, 2907}

\bibitem[\protect\citeauthoryear{{Schlegel}, {Finkbeiner}  \&
  {Davis}}{{Schlegel} et~al.}{1998}]{Schlegel_Finkbeiner.1998}
{Schlegel} D.~J.,  {Finkbeiner} D.~P.,   {Davis} M.,  1998, \mn@doi [\apj]
  {10.1086/305772}, \href {http://adsabs.harvard.edu/abs/1998ApJ...500..525S}
  {500, 525}

\bibitem[\protect\citeauthoryear{Seeger, Williams  \& Lawrence}{Seeger
  et~al.}{2003}]{Seeger_Williams.2003}
Seeger M.,  Williams C.,   Lawrence N.,  2003, in Artificial Intelligence and
  Statistics 9. No. EPFL-CONF-161318

\bibitem[\protect\citeauthoryear{Silverman}{Silverman}{1985}]{Silverman_only.1985}
Silverman B.~W.,  1985, Journal of the Royal Statistical Society. Series B
  (Methodological), pp 1--52

\bibitem[\protect\citeauthoryear{Snelson \& Ghahramani}{Snelson \&
  Ghahramani}{2006}]{Snelson_Ghahramani.2006}
Snelson E.,  Ghahramani Z.,  2006, in Advances in Neural Information Processing
  Systems. MIT press, pp 1257--1264

\bibitem[\protect\citeauthoryear{Snelson \& Ghahramani}{Snelson \&
  Ghahramani}{2007}]{Snelson_Ghahramani.2007}
Snelson E.,  Ghahramani Z.,  2007, in Proceedings of Artificial Intelligence
  and Statistics (AISTATS.

\bibitem[\protect\citeauthoryear{{Tierney}}{{Tierney}}{1994}]{Tierney_only.1994}
{Tierney} L.,  1994, Annals of Statistics, 22, 1701

\bibitem[\protect\citeauthoryear{Titsias}{Titsias}{2009}]{Titsias_only.2009}
Titsias M.~K.,  2009, in International Conference on Artificial Intelligence
  and Statistics. pp 567--574

\bibitem[\protect\citeauthoryear{Tresp}{Tresp}{2000}]{Tresp_only.2000}
Tresp V.,  2000, Neural Computation, 12, 2719

\bibitem[\protect\citeauthoryear{Vanhatalo, Pietil{\"a}inen  \&
  Vehtari}{Vanhatalo et~al.}{2010}]{Vanhatalo_Pietilainen.2010}
Vanhatalo J.,  Pietil{\"a}inen V.,   Vehtari A.,  2010, Statistics in medicine,
  29, 1580

\bibitem[\protect\citeauthoryear{{Vergely}, {Freire Ferrero}, {Siebert}  \&
  {Valette}}{{Vergely} et~al.}{2001}]{Vergely_FreireFerrero.2001}
{Vergely} J.-L.,  {Freire Ferrero} R.,  {Siebert} A.,   {Valette} B.,  2001,
  \mn@doi [\aap] {10.1051/0004-6361:20010006}, \href
  {http://adsabs.harvard.edu/abs/2001A%26A...366.1016V} {366, 1016}

\bibitem[\protect\citeauthoryear{Whittle}{Whittle}{1954}]{Whittle_only.1954}
Whittle P.,  1954, Biometrika, 41, pp. 434

\makeatother
\end{thebibliography}

\appendix

\section{Accelerating GP regression}\label{sec:accel}

As a consequence of their versatility, power and widespread adoption across a wide variety of fields there has been significant interest in extending the use of GP regression and inference to larger datasets. 
In this section we will briefly review some methods that accelerate GP regression.
We are not aware of any up-to-date and totally comprehensive review of such methods, however, \cite{Quinonero-Candela_Rasmussen.2005} and \cite{Rasmussen_Williams.2005} do examine a many of the methods discussed below in significantly more detail than we can afford them here.

The most trivial approach to accelerating GP regression
is to approximate the full solution by using only a small subset of
all the available data.
If we employ only $M$ of the original $N$ observations, such that $M
\ll N$, the CPU cost of the regression is reduced to $\mathcal{O}(M^3)$.
However, it is trivially apparent that this is far from optimal,
since, by its very nature, it requires discarding most available data.
So, it cannot hope to be a competitive solution.

An alternative simple approximation is to divide the entire region
studied into a number of small regions and then independently
implement GP regression in each, using only the
observations within each region.
This approximation is sometimes referred to as local Gaussian
processes.
If each local region contains $M$ observations, so that we again have
that $M \ll N$, the CPU cost is reduced to just
$\mathcal{O}(M^2N)$.
There are two significant drawbacks to this approach, however.
First, it will introduce discontinuities at the transitions between
regions.
In extinction mapping these will manifest as fingers of God.
More importantly, if the size of the regions is small compared to the scale of
the covariance kernel then there will be a very significant loss of
precision, because when estimating the value of the field at any given
point one is only using a subset of the data that carries relevant
information.

To overcome the problem of discontinuities inherent with local GPs, one could employ product of experts \citep[PoE][]{Hinton_only.2002}, a popular
machine learning technique that seeks to simplify complicated inference problems.
The concept behind PoE is that one approximates the true probability distribution by the product of a number of simple probability distributions.
However,  whilst a PoE approximation to GP regression can perform well in regions rich with observations, it produces estimates that are overconfident, particularly in regions where observations are sparse \citep{Deisenroth_Ng.2015}.

Bayesian committee machines (BCMs) were introduced by \cite{Tresp_only.2000} as a way of combining information from distinct datasets, with a particular focus on GP regression.
They are conceptually similar to PoE, but avoid that approach's problem with overconfidence.
However, they have a fundamental drawback in that they cannot deal with uncertainty on the hyperparameters: their entire derivation is conditioned on fixed hyperparameters.
Therefore, given that we expect the a priori unknown hyperparameters that describe the large-scale structure of extinction to have a key role in the determination of 3D maps, BCMs cannot directly enable the production of 3d extinction maps within our scheme.

There are a number of methods for GP regression that employ a variational Bayesian approach \citep[e.g][]{Titsias_only.2009, Hensman_Fusi.2013}.
However, these approaches generally assume that the positions of the observations are known exactly.
As this is not the case with our data, these methods are not so easily applicable here.

It has long been recognised \citep[e.g][]{Whittle_only.1954} that if one has a stationary covariance function and a regular grid of observations then the covariance matrix can be diagonalised through a Fourier transform.
The extension of this concept to irregularly grided data and non-stationary covariance functions is provided by the use of Stochastic Partial differential Equations \citep[SPDEs]{Lindgren_Rue.2011}.
By combining the use of SPDEs with the Integrated Nested Laplace Approximation \citep[INLA]{Rue_Martino.2009} \cite{Lindgren_Rue.2011} were able to quickly perform GP regression with large datasets and non-stationary covariance functions.
Two of the key stages in the application of SPDEs are the use of a coordinate transformation to shift to a space where the covariance function is stationary and the triangulation of the space such that all observations lie on the vertices of triangles (or tetrahedra in 3D).
However, in our case both the covariance function (which depends on the mean function) and the locations of the observations are unknown.
Consequently, not only would implementing an SPDE approach be rather complicated in our case, but also much of the CPU time saved by the use of SPDEs would be lost to repeatedly recalculating the coordinate transform and retriangularising the space.

All the above approaches are approximations to full GP
regression and so will inevitably involve some loss of precision.
An alternative is discussed by \cite{Ambikasaran_Foreman-Mackey.2016},
who recognise that the covariance matrices of GPs are
hierarchical off-diagonal low-rank (HODLR), enabling them to be
factorised hierarchically following \cite{Ambikasaran_Darve.2013}.
\cite{Ambikasaran_Foreman-Mackey.2016} show that one can then
factorise the covariance matrix in $\mathcal{O}(N \log^2 N)$
time.
Subsequently finding the inverse or determinant of the covariance
matrix is very fast and scales as $\mathcal{O}(N \log N)$.
In our application, however, the uncertainties in the mean extinction
function and the distances to individual stars mean that the
covariance function itself is unknown.
This would mean that we could not avoid repeatedly applying the
relatively expensive factorisation step.
Consequently, although this approach is significantly faster than the
naive application of GP regression, we found that it was
not competitive with the approach we describe in
section~\ref{sec:scheme}.

\subsection{Sparse Approximate GP Regression}\label{sec:sparse}

We start by noting that the joint, marginal and conditional distributions of two sets of points that are drawn from the same GP are all (multivariate) Gaussian.
We consider a GP with $N$ observations $\vy$ at locations $\vx$ that are partitioned into $K$ subsets, $\vy = \{\vy_1, \vy_2 \ldots \vy_K\}$ and $\vx = \{\vx_1, \vx_2 \ldots \vx_K\}$.
We further assume, for the sake of simplicity, that all of these subsets contain $M = N/K$ observations where $M<<N$.
We can then use the rules of conditional probability to decompose the GP probability
\begin{equation}
\pr (\vy | \vtheta) = \prod_{k=1}^K \pr(\vy_k | \vy_1, \ldots \vy_{k-1} , \vtheta) .
\label{eqn:exactfactorise}
\end{equation}
Thus, when conditioned on the first $k-1$ blocks, the $k^{\rm th}$ block follows a multivariate Gaussian PDF with the mean vector and covariance matrix
\begin{equation}
\begin{split}
&\vm_{k | 1 \ldots k-1} = \vm_k + \vSigma_{k,1 \ldots k-1} \vSigma_{1 \ldots k-1, 1 \ldots k-1}^{-1} (\vy_{1 \ldots k-1} - \vm_{1 \ldots k-1} ) \\
&\vSigma_{k | 1 \ldots k-1} = \vSigma_{k,k} - \vSigma_{k,1 \ldots k-1} \vSigma_{1 \ldots k-1, 1 \ldots k-1}^{-1} \vSigma_{1 \ldots k-1, k} .
\end{split}
\label{eqn:true_meancov}
\end{equation}
Where 
\begin{equation}
\begin{pmatrix}
\vy_{1 \ldots k-1}\\
\vy_k
\end{pmatrix} \sim \mathcal{N} \left(
\begin{pmatrix}
\vm_{1 \ldots k-1}\\
\vm_{k}
\end{pmatrix} ,
\begin{pmatrix}
\vSigma_{1 \ldots k-1, 1 \ldots k-1} & \vSigma_{1 \ldots k-1, k} \\
\vSigma_{k,1 \ldots k-1} & \vSigma_{k,k}
\end{pmatrix} \right),
\end{equation}
so that, for example, $\vm_{k}$ and $\vSigma_{k,k}$ are the marginal mean and covariance for the $k^{\rm th}$ block.
It is important to note that factorising the PDF in this manner does not reduce the CPU or memory costs involved in calculating or sampling from $\pr (\vy | \vtheta)$, which retain their $\mathcal{O}(N^3)$ and $\mathcal{O}(N^2)$ scaling respectively.

A simple approximation to the factorisation above would be to assume that off-diagonal blocks in the covariance matrix are populated exclusively with zeros, then
\begin{equation}
\begin{split}
&\pr (\vy | \vtheta) \approx \prod_{k=1}^K \pr(\vy_k | \vtheta) \\
&\vm_{k | 1 \ldots k-1} \approx \vm_k \\
&\vSigma_{k | 1 \ldots k-1} \approx \vSigma_{k,k} .\\
\end{split}
\end{equation}
This approximation is essentially the local GP approximation: when conditioned on the hyperparameters, the values of the GP in different blocks are independent.
This would result in significant and attractive computational savings: CPU time and memory size now scale as $\mathcal{O}(M^2 N)$ and $\mathcal{O}(MN)$.
However, in practice this is rarely a sensible approximation and will lead to large discontinuities between blocks and over-fitting of hyperparameters.

But, what if we could find a computationally cheap way to approximate the impact of the off-diagonal blocks in the covariance matrix?
There exists a family of approaches that seek to obtain a sparse approximation to the full covariance matrix using so-called `inducing points'.
We introduce the values $\vu$ of a GP at $R$ inducing points $\vx_u$ and note that, by the definition of a GP, both $\pr(\vy | \vtheta)$ and $\pr(\vy, \vu | \vtheta)$ follow multivariate Gaussian PDFs.
\cite{Snelson_Ghahramani.2007} propose the Partially Independent Training Conditional (PITC) approximation, whereby one assumes that subsets of observations are independent given the value of the GP at the inducing point, i.e.
\begin{equation}
\pr(\vy_k | \vy_1, \ldots \vy_{k-1} , \vu, \vtheta) \approx \pr(\vy_k | \vu, \vtheta)
\label{eqn:PITC_base}
\end{equation}
We can therefore approximate the factorised PDF above
\begin{equation}
\begin{split}
\pr (\vy | \vtheta) &= \int \, d\vu \pr(\vu | \vtheta)  \prod_{k=1}^K \pr(\vy_k | \vy_1, \ldots \vy_{k-1} , \vu, \vtheta) \\
&\approx \int \, d\vu \pr(\vu | \vtheta) \prod_{k=1}^K \pr(\vy_k | \vu, \vtheta) .
\end{split}
\end{equation}
This approximate likelihood can then be employed to obtain the posterior required in the learning phase of GP regression \eqref{eqn:example_learning}.
\begin{equation}
\begin{split}
\pr (\vtheta | \vy) &\approx \frac{\pr(\vtheta)}{\pr(\vy)} \int \, d\vu \pr(\vu | \vtheta) \prod_{k=1}^K \pr(\vy_k | \vu, \vtheta) \\
\pr (\vu, \vtheta| \vy) &\approx \frac{\pr(\vu, \vtheta)}{\pr(\vy)} \prod_{k=1}^K \pr(\vy_k | \vu, \vtheta) . 
\end{split}
\label{eqn:PITC_posterior}
\end{equation}
When conditioned on $\vu$, which has mean $m_{\vu}$ and covariance $\vSigma_{\vu,\vu}$, and given the PITC approximation, the means and covariances of the $k^{\rm th}$ block are
\begin{equation}
\begin{split}
&\vm_{k | \vu} \approx \vm_k +  \vSigma_{k, \vu} \vSigma_{\vu,\vu}^{-1} (\vu - m_{\vu} )\\
&\vSigma_{k | \vu} \approx \vSigma_{k,k} - \vSigma_{k,\vu} \vSigma_{\vu,\vu}^{-1} \vSigma_{\vu, k} .\\
\end{split}
\end{equation}
From which it is clear how the inducing points provide an approximation to the true mean and covariance \eqref{eqn:true_meancov}.

Having employed the PITC approximation to accelerate the learning phase we now turn to the prediction phase.
As discussed in section~\ref{sec:GPs}, we can use observations of a GP at some points to constrain it at others.
Specifically, the value of the GP at some unobserved position(s) is related to that at other observed positions,
\begin{equation}
\pr(\vy_{\star} | \vy) = \int \, \d \theta \pr(\vy_{\star} | \vy, \theta) \pr(\theta | \vy) ,
\end{equation}
PITC provides us with a computationally cheap approximation to $\pr(\theta | \vy)$.
However, we still need to calculate $\pr(\vy_{\star} | \vy, \theta)$.
In vanilla GP regression, if the covariances of $\vy$ given $\theta$ are stored during the learning phase, then the CPU time needed to find or sample from $\pr(\vy_{\star} | \vy, \theta)$ scales as $\mathcal{O}(N^2)$.
In most astrophysical applications this cost will easily become restrictive.

Instead, \cite{Snelson_Ghahramani.2007} further propose the partially independent conditional (PIC) approximation, that extends further upon PITC.
To illustrate this, we initially consider estimating the value $y_{\star}$ of a GP at a single test point $x_{\star}$.
If we break $\vy$ into blocks as in the PITC approximation, we obtain the factorisation
\begin{equation}
\pr(y_{\star}, \vy | \theta) = \pr(y_{\star} | \vy_1 \ldots \vy_K, \theta) \prod_{k=1}^K \pr(\vy_k | \vy_1 \ldots \vy_{k-1}, \theta) .
\end{equation}
As with the exact factorisation of $\pr(\vy | \theta)$~\eqref{eqn:exactfactorise}, this factorisation provides no computational benefit.
The PIC approximation assumes that the value $y_{\star}$ of the GP at the prediction point depends only on one block $\vy_i$ of $\vy$, when conditioned $\vu$.
\begin{equation}
\pr(y_{\star} | \vy_1 \ldots \vy_K, \vu, \theta) \approx \pr(y_{\star} | \vy_i, \vu, \theta) .
\end{equation}
Typically the block chosen will cover the region of space in which $x_{\star}$ falls.
Combining this with the existing approximation~\eqref{eqn:PITC_base} obtains
\begin{equation}
\pr(y_{\star}, \vy | \vu, \theta) \approx \pr(y_{\star} | \vy_i, \vu, \theta) \prod_{k=1}^K \pr(\vy_k | \vu, \theta) ,
\end{equation}
and so
\begin{equation}
\pr(y_{\star} | \vy, \vu, \theta) \approx \pr(y_{\star} | \vy_i, \vu, \theta) .
\end{equation}
Calculating or sampling from this is cheap, scaling as $\mathcal{O}(M^2)$ since it is only conditioned on a single block $\vy_i$, in addition to the inducing points and hyperparameters, rather than all $K$ blocks.
Therefore,
\begin{equation}
\pr(y_{\star} | \vy) \approx \iint \, \d\vu \, \d\theta \pr(y_{\star} | \vy_i, \vu, \theta) \pr (\vu, \vtheta| \vy),
\end{equation}
where we can employ the PITC approximation~\eqref{eqn:PITC_posterior} to $\pr (\vu, \vtheta| \vy)$.

In general, we wish to predict the values $\vy_{\star}$ of the GP at many prediction points $\vx_{\star}$.
We can break $\vy_{\star}$ into blocks analogously to $\vy$, typically based on the positions $\vx_{\star}$.
Then 
\begin{equation}
\pr(\vy_{\star} | \vy, \vu, \theta) \approx \prod_{k=1}^K \pr(\vy_{\star,k} | \vy_k, \vu, \theta) ,
\end{equation} 
therefore,
\begin{equation}
\pr(\vy_{\star} | \vy) \approx \iint \d\vu \, \d\theta \, \pr (\vu, \vtheta| \vy) \prod_{k=1}^K \pr(\vy_{\star,k} | \vy_k, \vu, \theta).
\end{equation}

The CPU time and memory requirements of the training phase of vanilla GP regression are dominated by the need to invert and store an $N \times N$ covariance matrix, resulting in them scaling as $\mathcal{O}(N^3)$ and $\mathcal{O}(N^2)$ respectively.
However, making the PI(T)C approximation of \cite{Snelson_Ghahramani.2007} alters our needs so that we instead must invert and store $K$ $M \times M$ covariance matrices.
As a result, CPU time and memory requirements instead scale as $\mathcal{O}(NM^2)$ and $\mathcal{O}(NM)$ respectively.
Given that $M << N$, this potentially results in massive savings.
In addition, the PIC approximation reduces the cost of the prediction phase, from $\mathcal{O}(N^2)$ under vanilla GP regression to $\mathcal{O}(NM)$.
Therefore, PIC potentially makes GP regression viable in a much wider range of problems than vanilla GP regression can approach.

However, whilst PIC can produce excellent results, it cannot be straightforwardly parellelised without significant inter-thread communication.
Consequently, the size of a region that can be easily studied in this way is in practice often limited to that that can be contained on a single computing node.

PI(T)C is just one of a number of inducing point methods, as reviewed by \cite{Quinonero-Candela_Rasmussen.2005}.
The other methods in this group include Subset of Regressors \citep{Silverman_only.1985}, Projected Latent Variables \citep{Seeger_Williams.2003} and Fully Independent Conditional \citep{Snelson_Ghahramani.2006}.
However, as \cite{Snelson_Ghahramani.2007} demonstrate, PI(T)C is the most sophisticated and generally most accurate of them.

\subsection{Expectation Propagation}\label{sec:EP}

Expectation propagation (EP) seeks to make inferences based on
large datasets, breaking the data down into smaller sets, which
are analysed separately, before combining the results together.
EP was originally derived by \cite{Minka_only.2001}, whilst
\cite{Bishop_only.2006} and \cite{Gelman_Vehtari.2015} provide accessible
introductions.

We are not aware of any prior use of EP in the astrophysics
literature, though \cite{Gelman_Vehtari.2015} do briefly examine a
highly simplified astronomy inspired example.
However, it has been widely applied in a variety of other fields,
including genetics \citep{Parts_Stegle.2011}, geostatistics
\citep{Cornford_Csato.2005} and in examining future employment
prospects \citep{Frey_Osborne.2013}.

The goal of expectation propagation is to approximate the posterior
\begin{equation}
\pr(\hyper | \vy) = \frac1{\pr(\vy)}\pr(\hyper)\prod_{k=1}^K \pr(\vy_k | \hyper),
\label{eqn:EPposterior}
\end{equation}
by a factorised distribution of the form
\begin{equation}
q(\hyper) =\pr(\hyper) \prod_{n=1}^K\frac1{Z_k} q_k(\hyper) ,  \label{eqn:EP}
\end{equation}
in which the (normalised) fitting functions $q_k$, also known as site distributions, have some simple
parametrized functional form.
The idea then is to adjust the parameters describing the $\{q_k\}$ to
optimise the agreement with the posterior~\eqref{eqn:EPposterior}.

The EP algorithm uses the Kullback--Leibler (KL) divergence,
\begin{equation}
  \hbox{KL}(f||g)\equiv
-\int f(\hyper)\log\left(\frac{g(\hyper)}{f(\hyper)}\right)\,\d\hyper,
\label{eqn:KLdefn}
\end{equation}
to quantify the differences between various functions $f(\hyper)$ and
$g(\hyper)$ related to the the posterior~\eqref{eqn:EPposterior} and
the factorised approximation~\eqref{eqn:EP}.
To begin, consider
\begin{equation}
  \hbox{KL}(p||q)=\hbox{KL}\left(
    \frac1{p(\vy)}\pr(\hyper)\prod_k\pr(\vy_k|\hyper)\bigg|\bigg|
    \frac1Z\pr(\hyper)\prod_kq_k(\hyper)
    \right),
\end{equation}
which would be very difficult to minimise directly.
The key idea of EP is to focus on just one factor $q_l$ at a time,
replacing all factors $\pr(\vy_k|\hyper)$ for $k\ne l$ in the
first argument by the corresponding~$q_k(\hyper)/Z_k$, namely
\begin{equation}
  \hbox{KL}\left(
    \frac1{p(\vy)}\pr(\hyper)\pr(\vy_l|\hyper)\prod_{k\ne l}q_k(\hyper)\bigg|\bigg|
    \frac1{Z_l}\pr(\hyper)q_l(\hyper)\prod_{k\ne l}q_k(\hyper)
    \right).
\label{eqn:KL1}
\end{equation}
So, each $q_l$ is fit just to the corresponding posterior factor
$\pr(\vy_l|\hyper)$, but weighted to include the other $q_{k\ne
  l}$ and the prior~$\pr(\hyper)$.
This is nicely visualised by Fig.~2 of \cite{Gelman_Vehtari.2015}.

Introducing the {\it cavity} distribution
\begin{equation}
  q_{\_ l}(\hyper)\equiv\frac1{q_l(\hyper)}q(\hyper)
  =\frac1{Z_l}\pr(\hyper)\prod_{k\ne l}q_k(\hyper)
\label{eqn:cavity}
\end{equation}
and the {\it tilted} distribution
\begin{equation}
  q_{\backslash l}(\hyper)\equiv
  \pr(\vy_l|\hyper)q_{\_ l}(\hyper),
\label{eqn:tilted}
\end{equation}
enables the KL divergence~\eqref{eqn:KL1} to be rewritten more compactly as
\begin{equation}
  \hbox{KL}\left(\frac1{C_l}
    q_{\backslash l}(\hyper)
    \bigg|\bigg|
    q_{\_ l}(\hyper)q_l(\hyper)
    \right), 
\label{eqn:KL1'}
\end{equation}
with $C_l\equiv\int q_{\backslash l}(\hyper)\d\hyper$.
If the form assumed for the $q_k(\vtheta)$ (and also for the prior
$\pr(\vtheta)$) belongs to the exponential family of distributions
then the second argument is itself another member of the exponential
family and the KL divergence is easy to minimise
\citep[e.g.,][]{Bishop_only.2006}: just take $Z_l=C_l$ and adjust the
parameters of $q_l(\vtheta)$ so that the sufficient statistics of
$\frac1{Z_l}q_{\backslash l}(\hyper)$ and $q_{\_
  l}(\hyper)q_l(\hyper)$ match.
In particular, if we adopt Gaussian functional forms for the $q_k$,
then the best choice of mean and covariance for the factor $q_l$ is
given by matching the first- and second-order moments of
$q_l(\hyper)q_{\_l}(\hyper)$ to those of $\frac1{Z_l}q_{\backslash
  l}(\hyper)$.

The key detail to appreciate from the above is the central roles the cavity and tilted distributions play.
In particular that we can define the problem of finding the best approximation $q(\hyper)$ to involve only the cavity, tilted and site distributions.
The EP algorithm then simply refines the site approximations $q_k(\hyper)$ so that the divergence~\eqref{eqn:KL1'} is minimised and the best approximation to the underlying distribution is found.

EP can be carried out with these updates to the $q_k$ performed in series or in parallel.
The parallel procedure can be summarised as follows \citep[paraphrasing][]{Gelman_Vehtari.2015}:
\begin{itemize}[leftmargin=*,labelindent=16pt]
\item \textbf{Initialise:} Partition the data into $K$ groups, so that
  $\vy = \{\vy_1 \ldots \vy_K\}$. Then approximate each
  $\pr(\vy_k | \hyper)$ with $q_k(\hyper)$ chosen from a simple
  distribution (e.g. Gaussian).
  Use  \eqref{eqn:EP} to construct a first guess to $q(\hyper)$.
\item \textbf{Iterate:} Perform the following until convergence is achieved:
\begin{itemize}[leftmargin=*,labelindent=16pt]
\item For $k = 1 \ldots K$:
\begin{itemize}[leftmargin=*,labelindent=16pt]
\item Construct the cavity and tilted distributions,
  $q_{\_ k}(\hyper) = q(\hyper) / q_k(\hyper)$ and $q_{\backslash k}(\hyper)
  = \pr(\vy_k|\hyper) q_{\_ k}(\hyper)$.  Generally the latter
  will not be available analytically, but one can sample
  from it using, e.g., MCMC.
\item Find a new approximation $q_k'(\hyper)/Z_k'$ that minimises~\eqref{eqn:KL1'}
  by, e.g., moment matching.
\end{itemize}
\item Construct the updated approximation $q'(\hyper)=\pr(\hyper) \prod_{k=1}^n\frac1{Z'_k}q_k'(\hyper)$ and adopt this as the new approximation $q(\hyper)$.
\item Check for convergence.  If reached then terminate, so that the final approximation is given by $q(\hyper)$
\end{itemize}
\end{itemize}
The extension to hierarchical models, such as the one we employ, is straightforward and described by \cite{Gelman_Vehtari.2015}.
An advantage of using EP to study hierarchical models is that it allows `local' parameters to be partitioned, such that they only enter one of the likelihood terms.
The only practical difference compared to the non-hierarchical algorithm outlined above occurs when calculating the tilted distribution.
If there are additional `local' parameters $\alpha_k$ that sit between $\hyper$ and the data, then the \textit{local unmarginalised} tilted distribution $q_{\backslash k}(\alpha_k,\hyper) = \pr(\vy_k|\alpha_k,\hyper) \pr(\alpha_k|\hyper) q_{\_ k}(\hyper)$ is found for each  $k = 1 \ldots K$.
As before, one typically samples from this distribution numerically.
Then the marginal tilted distribution $q_{\backslash k}(\hyper)$ is found by marginalising $\alpha_k$.
In some cases this marginalisation can be performed analytically.
However, more typically the marginalisation is performed numerically: if the local unmarginalised tilted distribution is found by MCMC then the marginalisation is trivial.

In some respects it is possible to think of EP as a generalisation of BCMs.
Specifically, BCMs are equivalent to a single pass of a parallel EP algorithm, conditioned on fixed hyperparameters.
However, in contrast to BCMs, expectation propagation is suitable for use with hierarchical models with non-fixed hyperparameters and so is a good fit for our purposes.

\section{Details of the moment matching in EP}
\label{sec:momentmatching}

Since $\va_k$ and $\vu$ are drawn from a GP, it follows that $\pr(\va_k, \vu | \macromodel)$ is Gaussian.  We express this Gaussian as
\begin{equation}
\va_k, \vu | \macromodel \sim \mathcal{N} \left(
\begin{pmatrix}
\mu_k \\
m
\end{pmatrix} 
,
\begin{pmatrix}
\Sigma_k & \rho_k \\
\rho_k^T & C
\end{pmatrix} 
\right) .
\end{equation}
Similarly, 
\begin{equation}
\vatilde_i, \vu | \hyper \sim \mathcal{N} \left(
\begin{pmatrix}
\mu_i \\
m
\end{pmatrix} 
,
\begin{pmatrix}
\Sigma_i + \sigma_i & \rho_i \\
\rho_i^T & C
\end{pmatrix} 
\right) .
\end{equation}
The reduced cavity distribution is a Gaussian, which we may write as
\begin{equation}
q'_{\_ k}(\vu, \macromodel) \sim \mathcal{N} \left(
\begin{pmatrix}
m_{\_ k} \\
\psi_{\_ k}
\end{pmatrix} 
,
\begin{pmatrix}
C_{\_ k} & \rho_{\_ k} \\
\rho_{\_ k}^T & \Omega_{\_ k}
\end{pmatrix} 
\right) .
\end{equation}
With these expressions $q'_k(\vu|\macromodel)$ and $\pr(\vu|\vatilde_k,\macromodel)$ are given by
\begin{equation}
\begin{split}
 \vu | \vatilde_k, \macromodel &\sim \mathcal{N} \left(m'_k, C'_k \right) \\
 q'_{\_ k}(\vu | \macromodel) &\sim \mathcal{N} \left(m'_{\_ k}, C'_{\_ k} \right) ,
\end{split}
\end{equation}
where
\begin{equation}
\begin{split}
m'_k &= m + \rho_k \left(\Sigma_k + \sigma_k \right)^{-1} (\vatilde_k-\mu_k), \\
C'_k &= C + \rho_k \left(\Sigma_k + \sigma_k \right)^{-1} \rho_n^T, \\
m'_{\_ k} &= m_{\_ k} + \rho_{\_ k} \Omega_{\_ k}^{-1} (\macromodel - \psi_{\_ k}), \\
C'_{\_ k} &= C_{\_ k} + \rho_{\_ k} \Omega_{\_ k}^{-1}\rho_{\_ k}^T .\\
\end{split}
\end{equation}
The Gaussian $\overline{q}_k(\vu | \vatilde_k, \macromodel)$ and its normalising factor $\tiltednormalizer_k$ are given by
\begin{equation}
\begin{split}
&\overline{q}_k(\vu | \vatilde_k, \macromodel) \sim \mathcal{N} \left(\overline{m}_k, \overline{C}_k \right) \\
&\overline{C}_k^{-1} = C_k^{\prime-1} + C_{\_k}^{\prime -1}, \\
&\overline{C}_k^{-1} \overline{m}_k = C_k^{\prime -1} m'_k + C_{\_k}^{\prime -1} m'_{\_k}, \\
&\tiltednormalizer_k(\vatilde_k, \macromodel) = \frac{|\overline{C}_k|^{\frac{1}{2}}}{|C'_k |^{\frac{1}{2}} |C'_{\_k}|^{\frac{1}{2} }}
\times\\
&\quad\exp \left[ -\frac{1}{2} \left ( m_k^{\prime T} C_k^{\prime -1} m'_k + m_{\_k}^{\prime T} C_{\_k}^{\prime -1} m'_{\_k} - \overline{m}_k^T \overline{C}_k^{-1} \overline{m}_k \right) \right]. \\
\end{split}
\end{equation}

\section{Derivation of the approximate prediction distributions}\label{app:prediction}

We start by recalling that the 3d map of extinction formally follows the posterior predictive distribution~\eqref{eqn:map_dist}.
Following some manipulation, not repeated here for the sake of brevity, we can rearrange this to a form similar that of the simpler example that was discussed in section~\ref{sec:simplified},
\begin{equation}
\begin{split}
& \pr(\va_\star | \vytilde, \vx_\star, \vltilde,\vbtilde,\micromodel,\galaxymodel)\\
&\quad= \int \d\macromodel \, \sum_{\vz} \int \d\vs \, \Big( \pr(\vs,\vz|\macromodel,\vytilde,\vltilde,\vbtilde,\micromodel,\galaxymodel) \pr(\macromodel|\vytilde,\vltilde,\vbtilde,\micromodel,\galaxymodel)
\\
&\qquad\qquad\qquad\qquad\times \int \d\va \, \pr(\va,\va_\star|\vs,\vz,\macromodel,\vytilde,\vltilde,\vbtilde,\micromodel,\galaxymodel) \Big) .
\end{split}
\end{equation}
Now we introduce $\vu$ into the integrand and marginalise, 
\begin{equation}
\begin{split}
& \pr(\va_\star | \vytilde, \vx_\star, \vltilde,\vbtilde,\micromodel,\galaxymodel)\\
&\quad= \iint \d\vu \, \d\macromodel \, \sum_{\vz} \int \d\vs \, \Big( \pr(\vs,\vz|\vu,\macromodel,\vytilde,\vltilde,\vbtilde,\micromodel,\galaxymodel) \pr(\vu,\macromodel|\vytilde,\vltilde,\vbtilde,\micromodel,\galaxymodel)
\\
&\qquad\qquad\qquad\qquad\times \int \d\va \, \pr(\va,\va_\star|\vs,\vz,\vu,\macromodel,\vytilde,\vltilde,\vbtilde,\micromodel,\galaxymodel) \Big) .
\end{split}
\end{equation}
By the use of the PIC approximation, we can obtain an approximate factorisation of the integrand,
\begin{equation}
\begin{split}
& \pr(\va_\star | \vytilde, \vx_\star, \vltilde,\vbtilde,\micromodel,\galaxymodel) \\
&\quad\approx \iint \d\vu \, \d\macromodel \, \Bigg( \pr(\vu,\macromodel|\vytilde,\vltilde,\vbtilde,\micromodel,\galaxymodel) \\
&\qquad \times \prod_{k=1}^K  \bigg[ \sum_{\vz_k} \int \d\vs_k \, \Big( \pr(\vs_k,\vz_k|\vu,\macromodel,\vytilde_k,\vltilde_k,\vbtilde_k,\micromodel,\galaxymodel) \\
&\qquad\qquad\qquad \int \d\va_k \, \pr(\va_k,\va_{\star,k}|\vs_k,\vz_k,\vu,\macromodel,\vytilde_k,\vltilde_k,\vbtilde_k,\micromodel,\galaxymodel) \Big) \bigg] \Bigg),
\end{split}
\end{equation}
were we have split $\va_\star$ into blocks corresponding to the sub-catalogues.
We can then use the EP approximation $q(\vu,\macromodel)$ to the posterior $\pr(\vu,\macromodel|\vytilde,\vltilde,\vbtilde,\micromodel,\galaxymodel)$, as found in the learning phase.
In addition, as we employ the GMM approximation to $XX$ we can replace $\vytilde$ by its sufficient statistics $\{\vatilde, \vstilde\}$ and perform the marginalisation
\begin{equation}
\begin{split}
&\int \d\va_k \, \pr(\va_k,\va_\star|\vs_k,\vz_k,\vu,\macromodel,\vatilde_k,\vltilde_k,\vbtilde_k,\micromodel,\galaxymodel) \\
&\quad = \pr(\va_\star|\vs_k,\vz_k,\vu,\macromodel,\vatilde_k,\vltilde_k,\vbtilde_k,\micromodel,\galaxymodel) 
\end{split}
\end{equation}
analytically.
We then obtain 
\begin{equation}
\begin{split}
& \pr(\va_\star | \vytilde, \vx_\star, \vltilde,\vbtilde,\micromodel,\galaxymodel) \\
&\quad\approx \iint \d\vu \, \d\macromodel \, \Bigg( q(\vu,\macromodel) \\
&\qquad \times \prod_{k=1}^K  \bigg[ \sum_{\vz_k} \int \d\vs_k \, \Big( \pr(\vs_k,\vz_k|\vu,\macromodel,\vstilde_k,\vltilde_k,\vbtilde_k,\micromodel,\galaxymodel) \\
&\qquad\qquad\qquad  \pr(\va_{\star,k}|\vs_k,\vz_k,\vu,\macromodel,\vatilde_k,\vltilde_k,\vbtilde_k,\micromodel,\galaxymodel) \Big) \bigg] \Bigg).
\end{split}
\end{equation}
This appears to be a formidable equation to deal with.
However, we note that if the EP algorithm has converged then the EP approximation and tilted distributions should be approximately equal, thus
\begin{equation}
\begin{split}
q(\vu,\macromodel) &\approx q_{\backslash k}(\vu,\macromodel) \\
q(\vu,\macromodel) \pr(\vs_k,\vz_k|\vu,\macromodel,\vstilde_k,\vltilde_k,\vbtilde_k,\micromodel,\galaxymodel) &\approx q_{\backslash k}(\vs_k, \vz_k,\vu,\macromodel),
\end{split}
\end{equation}
where $q_{\backslash k}(\vs_k, \vz_k,\vu,\macromodel)$ is the local unmarginalised tilted distribution.
Therefore,
\begin{equation}
\begin{split}
& \pr(\va_{\star,k} | \vytilde, \vx_\star, \vltilde,\vbtilde,\micromodel,\galaxymodel) \\
&\quad\approx \iint \d\vu \, \d\macromodel \, \sum_{\vz_k} \int \d\vs_k \,  \Big( q_{\backslash k}(\vs_k, \vz_k,\vu,\macromodel) \\
&\qquad\qquad\qquad  \pr(\va_{\star,k}|\vs_k,\vz_k,\vu,\macromodel,\vatilde_k,\vltilde_k,\vbtilde_k,\micromodel,\galaxymodel) \Big).
\end{split}
\end{equation}

\end{document}